\documentclass{aa} 

\usepackage{placeins}
\usepackage{graphicx}
\usepackage{graphicx,xcolor}
\usepackage{txfonts}
\usepackage{multirow}
\usepackage{lscape}
\usepackage{longtable}
\usepackage{comment}
\usepackage{caption}
\usepackage{footnote}
\makesavenoteenv{figure} 
\usepackage[breaklinks=true]{hyperref}
\pdfoutput=1
\usepackage{amsmath}

\usepackage{upgreek}

\newcommand{\Hii}{\ion{H}{II} }
\newcommand{\Hmol}{\mbox{H$_{\rm 2}$}}

\newcommand{\mum}{$\upmu$m}

\newcommand{\cc}{\mbox{cm$^{-3}$}}

\usepackage{savesym}
\usepackage{ulem} 
\usepackage[colorinlistoftodos]{todonotes} 
\usepackage[singlelinecheck=off]{caption}
\savesymbol{tablenum}
\usepackage[separate-uncertainty=true]{siunitx}
\restoresymbol{SIX}{tablenum}   
\usepackage{gensymb}
\usepackage{threeparttable}
\usepackage{bm}
\usepackage{natbib}

\DeclareSIUnit\mag{mag}

\DeclareSIUnit\angstrom{\text{\AA}}
\DeclareSIUnit\erg{\text{erg}}
\DeclareSIUnit\parsec{pc}

\begin{document}
\title{JWST observations of photodissociation regions\\ II. Warm molecular Hydrogen spectroscopy in the Horsehead nebula}
\author{	M. Zannese\inst{\ref{ias}}  \and
	P. Guillard\inst{\ref{iap}}\and
	A.~Abergel\inst{\ref{ias}}\and 
	E.~Habart\inst{\ref{ias}}  \and 
	P.~Dell'Ova\inst{\ref{ias}} \and 
	B.~Trahin\inst{\ref{stsci}} \and
	J.~Le~Bourlot\inst{\ref{obs},\ref{upc}}    \and
	K.~Misselt \inst{\ref{az}} \and
	D.~Van~De~Putte\inst{\ref{stsci}, \ref{uwo}}  \and 
	A.~N.~Witt \inst{\ref{toledo}}  \and 
	K.~D.~Gordon\inst{\ref{stsci},\ref{gent}}  \and 
	A.~Noriega-Crespo\inst{\ref{stsci}}   \and 
	M.~Baes\inst{\ref{gent}}  \and 
	P.~Bouchet\inst{\ref{cea}}  \and 
	B.~R.~Brandl\inst{\ref{leiden}}  \and 
	M.~Elyajouri\inst{\ref{ias}}  \and 
	O.~Kannavou\inst{\ref{ias}}  \and   
	P.~Klassen\inst{\ref{atc}}  \and 
	N.~Ysard\inst{\ref{irap},\ref{ias}}  
}		
\institute{
	Université Paris-Saclay, CNRS, Institut d'Astrophysique Spatiale, 91405 Orsay, France \label{ias}\\
	\email{marion.zannese@universite-paris-saclay.fr}  \and Sorbonne Universit\'{e}, CNRS, Institut d'Astrophysique de Paris, 98\,bis bd Arago, 75014 Paris, France \label{iap} \and
	Space Telescope Science Institute, 3700 San Martin Drive, Baltimore, MD, 21218, USA \label{stsci}  \and
	LUX, Observatoire de Paris, Universit\'e PSL, Sorbonne Universit\'e, CNRS, 92190 Meudon, France \label{obs}
	\and
	Universit\'e Paris-Cit\'e, Paris, France
	\label{upc} \and Steward Observatory, University of Arizona, Tucson, AZ 85721-0065, USA \label{az}
	\and Department of Physics \& Astronomy, The University of Western Ontario, London ON N6A 3K7 Canada \label{uwo} 
	\and
	Ritter Astrophysical Research Center, University of Toledo, Toledo, OH 43606, USA \label{toledo} \and
	%Ritter Astrophysical Research Center, University of Toledo, Toledo, OH 43606, USA \label{toledo}  \and
	Sterrenkundig Observatorium, Universiteit Gent, Krijgslaan 281 S9, B-9000 Gent, Belgium \label{gent} \and
	Universit\'e Paris-Saclay, Universit\'e Paris Cit\'e, CEA, CNRS, AIM, 91191 Gif-sur-Yvette, France \label{cea} \and
	Leiden Observatory, Leiden University, P.O. Box 9513, 2300 RA Leiden, The Netherlands \label{leiden} \and
	Faculty of Aerospace Engineering, Delft University of Technology, Kluyverweg 1, 2629 HS Delft, The Netherlands \label{fae}\and
	UK Astronomy Technology Centre, Royal Observatory Edinburgh, Blackford Hill, Edinburgh EH9 3HJ, UK \label{atc} \and
	Institut de Recherche en Astrophysique et Plan\'etologie, Universit\'e Toulouse III - Paul Sabatier, CNRS, CNES, 9 Av. du colonel Roche, 31028 Toulouse, France \label{irap} 
}

\date{Received ; accepted }

\abstract 
	{\Hmol\ is the most abundant molecule in the interstellar medium. Due to its excited form in irradiated regions, it is a useful tool to study photodissociation regions (PDRs), where radiative feedback from massive stars on molecular clouds is dominant. The \textit{James Webb Space Telescope} (JWST), with its high spatial resolution, sensitivity, and wavelength coverage provides unique access to the detection of most of \Hmol\ rotational and rovibrational lines and the analysis of its spatial morphology. %, which allows the probe of the spatial variation of physical parameters throughout the PDR at very small scales. 
    } %leave it empty if necessary  
	{Our goal is to use \Hmol\ line emission detected with the JWST in the Horsehead nebula to constrain the physical parameters (e.g., extinction, gas temperature, thermal pressure) throughout the PDR and its geometry.}
	{We use spectro-imaging data acquired using both the NIRSpec and MIRI-MRS instruments on board JWST to study the \Hmol\ spatial distribution at very small scales (down to 0.1"). The \Hmol\ line ratios allow us to constrain the extinction throughout the PDR. We then study in detail the excitation of \Hmol\ levels and use their analysis to derive physical parameters.}
	{We detect hundreds of \Hmol\ rotational and rovibrational lines in the Horsehead nebula. The study of \Hmol\ morphology reveals a spatial separation between \Hmol\ lines ($\sim$ 0.5") accross the PDR interface. FUV-pumped lines ($v=0$ $J_u>6$, $v>0$) peak closer to the edge of the PDR than thermalized lines. From \Hmol\ lines from the same upper level, we estimate the value of extinction throughout the PDR. We find that $A_V$ is increasing from the edge of the PDR to the second and third \Hmol\ filaments. We find $A_V = 0.3 \pm 1.3$ in the first filament and $A_V = 6.1 \pm 1.4$ in the second and third filaments. Then, we study the \Hmol\ excitation in different regions across the PDR. The excitation diagrams can be fitted by two excitation temperatures. As the first levels of \Hmol\ are thermalized, the colder temperature corresponds to the gas temperature. The second and hotter component corresponds to the FUV-pumped levels. In each filament, we then derive a gas temperature around $T \sim 500$ K. The temperature profile shows that the observed gas temperature is quite constant throughout the PDR, with a slight decline in each of the dissociation fronts. The study of the spatial distribution of \Hmol\ reveals that the column density of \Hmol\ is mostly located in the second and third filaments. Indeed, the column density estimated in the first filament is around $N(\Hmol) = (3.8 \pm 0.8) \times 10^{19}$ cm$^{-2}$ and $N(\Hmol) = (1.9 \pm 0.4) \times 10^{20}$ cm$^{-2}$ in the second and third filament, or five times higher. This study also reveals that the ortho-to-para ratio (OPR) is far from equilibrium and varies from OPR $\sim$ 2-2.5 at the edge of each dissociation front to a smaller value around OPR $\sim$ 1.3-1.5 deeper into the PDR. We observe a clear spatial separation of para and ortho rovibrational levels, as well as 0--0 S(2) and 0--0 S(1), indicating that efficient ortho-para conversion and preferential ortho self-shielding are driving the spatial variations of the OPR. Finally, we derive a thermal pressure in the first filament around $P_{\rm gas} \geq 6 \times 10^6$ K \cc, about ten times higher than that of the ionized gas. We highlight that template stationary 1D PDR models cannot account for the intrinsic 2D structure and the very high temperature observed in the Horsehead nebula. We argue the highly-excited, over-pressurized \Hmol\ gas at the edge of the PDR interface could originate from the mixing between the cold and hot phase induced by the photo-evaporation of the cloud.}
	{The analysis of \Hmol\ lines detected with JWST provides a unique access to the geometry and physical conditions in the Horsehead nebula at very small scale and reveals, for the first time, the possible importance of dynamical effects at the edge of the PDR. This study highlights, however, the need for extended modeling of these dynamical effects.} 

\keywords{Infrared: ISM: Horsehead, dust, extinction, molecules, lines and bands, photon-dominated region (PDR), H\,II regions, Techniques: photometric, spectroscopy, Methods: observational, data analysis}

 \titlerunning{JWST \Hmol\ spectroscopy of the Horsehead nebula}

 \maketitle
%-------------------------------------------------------------------

\section{Introduction}

Photodissociation Regions (PDRs) reprocess a significant part of the radiation output of young stars by re-emitting this energy in the infrared-millimeter wavelength regions through gas lines, aromatic bands, and thermal dust emission. The infrared emission from PDRs dominates the spectrum of galaxies and traces the regions where the radiative feedback is dominant. This is one of the major mechanisms to limit star formation \citep[e.g.,][]{inoguchi_factories_2020} by contributing to the dispersal of the cloud due to gas heating and angular-momentum addition. Moreover, the intense stellar Far-ultraviolet (FUV) radiation incident on PDRs plays a dominant role in the physics and chemistry of gas and dust \citep[for a review, see for example,][]{hollenbach_photodissociation_1999,wolfire_photodissociation_2022}. The study of these regions is thus essential for a better understanding of star formation and the evolution of interstellar matter. Moderately excited PDRs, such as the Horsehead nebula, are representative of most of the UV-illuminated molecular gas in the Milky Way and star-forming galaxies. The proximity and almost edge-on geometry of the Horsehead allow us to study in detail the physical structures of PDRs and the evolution of the physico-chemical characteristics of the gas and dust. The Horsehead is located on the western side of the molecular cloud Orion B at a distance of $\sim$ 400 pc \citep{anthony-twarog_h-beta_1982}. The Horsehead nebula emerges from the edge of the L1630 molecular complex and is seen as a dark cloud in silhouette against the \Hii region IC434 \citep[e.g.,][]{de_boer_diffuse_1983,neckel_spectroscopic_1985,compiegne_aromatic_2007,bally_kinematics_2018}. The Horsehead nebula is illuminated by the O9.5V binary system $\sigma$ Orionis \citep{warren_photometric_1977}, most likely on its backside as it is seen in silhouette. This system has an effective temperature of $T_{\rm eff} \sim 34600$ K \citep{schaerer_combined_1997} and is located at a projected distance of $\sim$ 3.5 pc from the edge of the Horsehead PDR. The incident UV field on the PDR is estimated to be $G_0 \sim 100$ \citep[with $G_0 = 1$ corresponding to a flux integrated between 91.2 and 240 nm of $1.6 \times 10^{-3}$ erg cm$^{-2}$ s$^{-1}$,][]{habing_interstellar_1968}.

H$_2$ is the most abundant molecule in galaxies. It is formed on the surface of interstellar grains, where the grains act as catalysts \citep{habart_empirical_2004,bron_surface_2014,wakelam_h_2017}. During its formation process, it is assumed, considering equipartition, that a third of the energy released from the reaction is converted into internal energy of the produced H$_2$. The other two-thirds of the H$_2$ formation energy is distributed between grain excitation and the kinetic energy of the released molecules. However, the branching ratio is unknown, and the distribution is probably uneven and depends on conditions in the PDR and the nature of the grains. In addition, in these regions, the first levels of \Hmol\ are excited by collisions due to the high densities, and the highly excited levels are populated by FUV pumping due to the intense UV field. \Hmol\ is a very useful tool for studying PDRs. Indeed, its first levels being thermalized at densities larger than $\approx 10^4$~\cc, \Hmol\ acts as a thermometer of the medium. Multiple lines of \Hmol\ have already been observed in the Horsehead nebula. The pure rotational lines detected with \textit{Spitzer} have revealed that only the levels $v=0, J<5$ are thermalized, while the more highly excited levels are mostly pumped by the UV field \citep{habart_excitation_2011}. Thus, the line intensities of these higher levels do not reflect the gas temperature but the fraction of the FUV photon flux pumping \Hmol. The observation of the 1 -- 0 S(1) line with the \textit{New Technology Telescope} (NTT) has revealed bright and narrow filaments at the illuminated edge of the PDR \citep{habart_density_2005}. The comparison between observations and PDR models has shown that there is a steep density gradient at the edge, with a scale length of $\le$0.02 pc (or $\sim 10"$) and $n_{\rm H} \sim 10^4$ \cc\ and $n_{\rm H} \sim 10^5$ \cc\ in the \Hmol\ emitting and inner molecular layers, respectively. \cite{habart_density_2005} have also shown that the Horsehead is viewed with a small inclination around $\sim 6$\degree, and thus it is not exactly edge-on. Finally, one of the striking results of the previous \Hmol\ data is that the observed column densities of rotationally excited \Hmol\ observed is much higher than PDR model predictions \citep{habart_excitation_2011}. In moderate excited PDRs such as the Horsehead, the discrepancy between the model and the data is about one order of magnitude for rotational levels $J_u\ge 5$. This disagreement suggests that our understanding of the formation and excitation of \Hmol\ and/or of PDRs heating and/or PDR dynamics is still incomplete. To make progress on this problem, JWST observations are essential because they allow spatial resolution of the smallest spatial scales and detection of many \Hmol\ rotational lines (with \textit{Spitzer} the detection was limited to few lines).

Several other studies have observed the Horsehead nebula throughout the years. For instance, observations of CO $J=1-0$ have allowed an estimate of the mean density of the Horsehead nebula \citep[$n_{\rm H} \sim 5 \times 10^3$ \cc,][]{pound_looking_2003}. More recently, observations of millimetric molecular emission with the Atacama Large Millimeter / Submillimeter Array (ALMA) at a high angular resolution of $\sim$ 0.5" have revealed a very thin atomic zone, with a size < 650 au, suggesting a very sharp transition between molecular and ionized gas \citep{hernandez-vera_extremely_2023}. From CO observations, they also derive the local gas density ($n_{\rm H} = (3.9 - 6.6) \times 10^4$ \cc) and the thermal pressure ($P_{\rm th} = (2.3 - 4.0) \times 10^6$ K \cc) at a distance of 15" from the edge of the cloud, defined as the ionization front.

\begin{figure*}

  \includegraphics[width=0.498\linewidth]{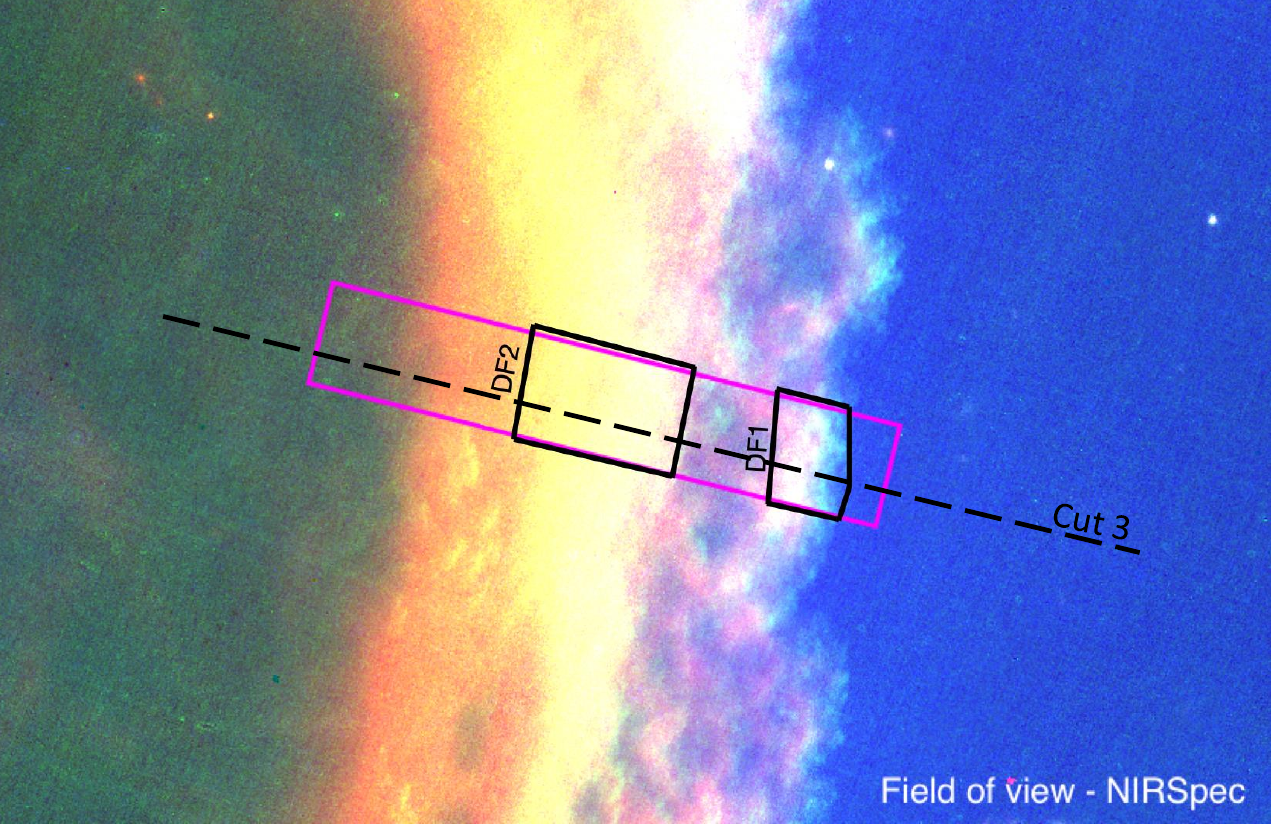}
  \includegraphics[width=0.498\linewidth]{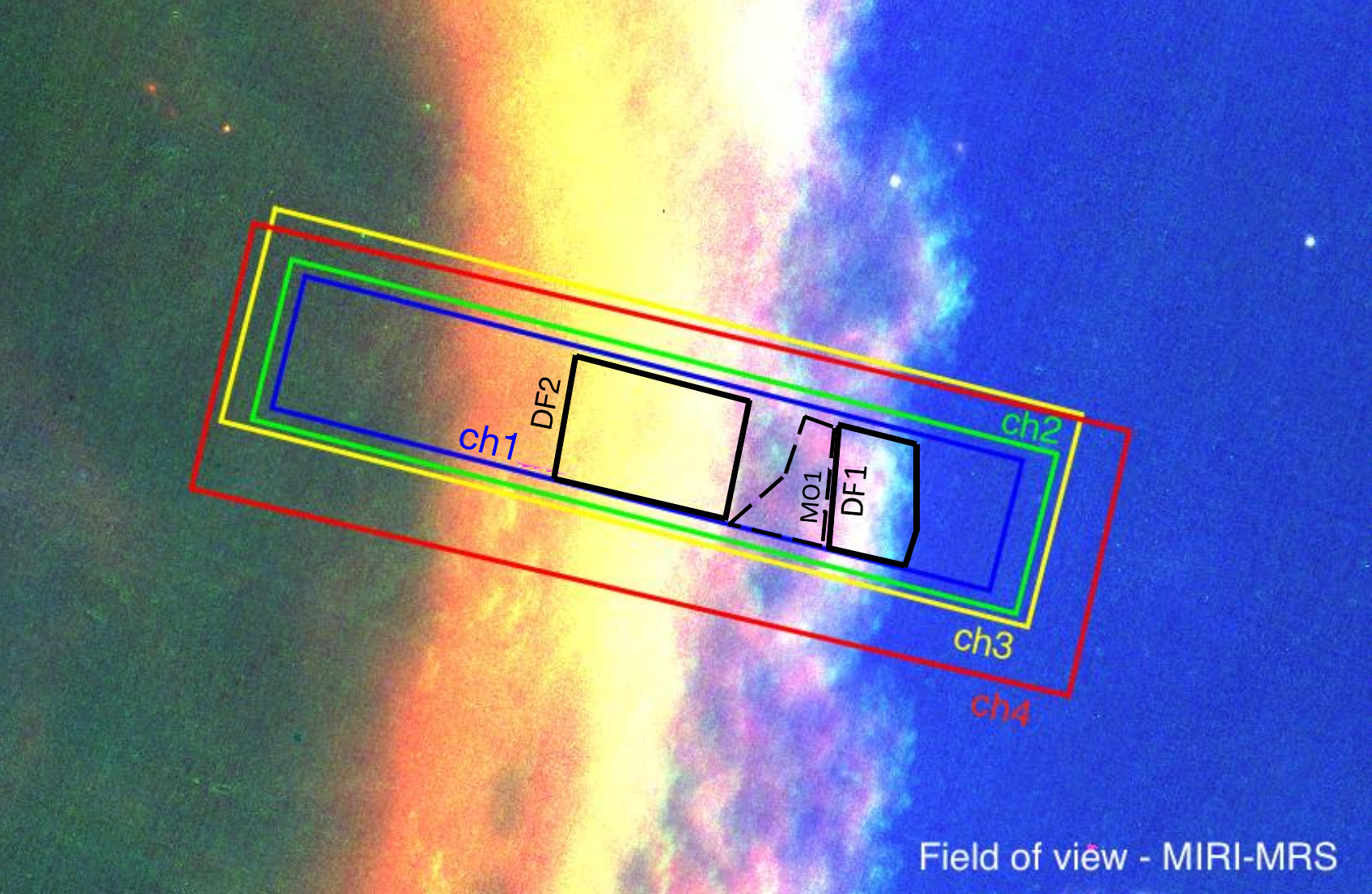}
  \includegraphics[width=\linewidth]{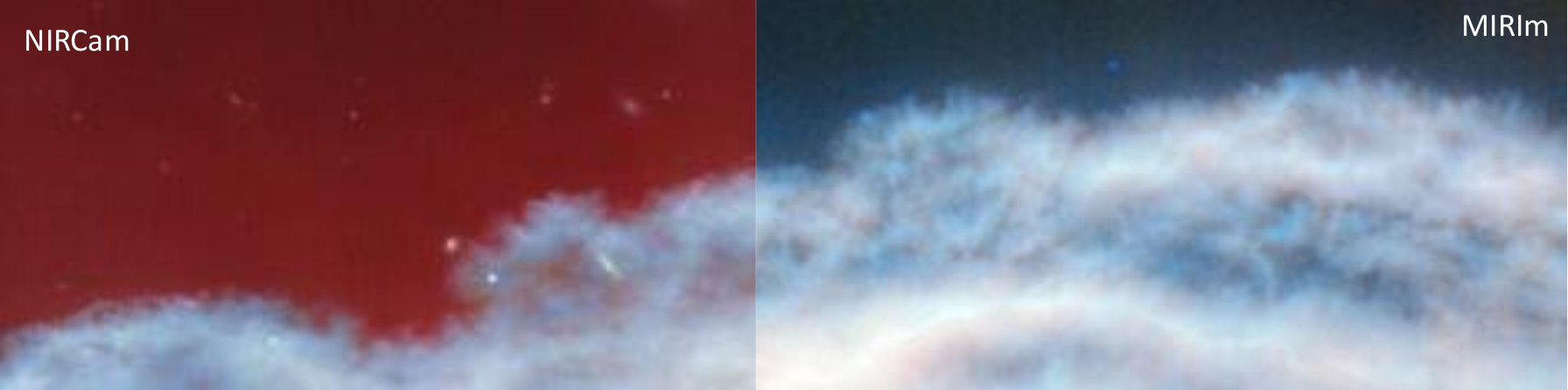}
 \caption{(Top) JWST NIRCam RGB image of the Horsehead nebula, located in the Orion molecular cloud. Red is the 3.35 $\upmu$m emission (F335M NIRCam filter), blue is the emission of Pa$\alpha$ (F187N filter), and green is the emission of the \Hmol\ 1--0 S(1) line at 2.12 $\upmu$m (F212N filter). (Left) The field of view of NIRSpec is overlayed in magenta on the image. (Right) The field of view of the different channels of MIRI-MRS is overlaid on the image (channel 1: blue, channel 2: green, channel 3: yellow, channel 4: red). The black boxes correspond to the aperture used to derive the spectra in the dissociation front regions and the black dashed boxes correspond to the aperture in the "molecular" region behind DF1 defined in \cite{misselt_jwst_2025}. The dashed line corresponds to the position of cut \#3 from \cite{abergel_jwst_2024}. (Bottom) JWST NIRCam and MIRI-MRS composite image of the Horsehead nebula, zoomed on the edge where the faint striated features, attributed to an evaporative flow, are more visible. This image is rotated by 90$\degree$ with respect to the top panel. Credit: ESA/Webb, NASA, CSA, K. Misselt (University of Arizona) and A. Abergel (IAS/University Paris-Saclay, CNRS).}
\label{fig:footpints_MRS}
\end{figure*}

The \textit{James Webb Space Telescope} (JWST), with its high sensitivity and high spatial resolution, gives access to the emission of \Hmol\ at very small scales (up to 0.1"), allowing a better understanding of the Horsehead nebula morphology than previous observations. The observations presented in this paper are part of the JWST Guaranteed Time Observations (GTO) program with ID number \#1192 (P.I.: Misselt) (see Fig. \ref{fig:footpints_MRS}). Imaging data analysis is presented in \cite{abergel_jwst_2024}. A network of faint striated features extending perpendicularly to the PDR front into the \Hii region has been revealed in the broadband NIRCam filter at 3.35\,\mum. This detection may indicate an entrainment of nano-dust particles in the evaporative flow. Maps of the 1 -- 0 S(1) line of \Hmol\ made by NIRCam reveal numerous sharp sub-structures on scales as small as 1.5". Similarly to the results of \cite{hernandez-vera_extremely_2023}, the imaging data reveal a very small size of the neutral atomic layer ($< 100$~au). The analysis of imaging in the broadband filters also shows strong color variations between the illuminated edge and the internal regions, which can be explained by dust attenuation if the Horsehead is illuminated from behind. It appears that dust attenuation is non-negligible over the entire spectral range of the JWST. Spectroscopic data analysis, presented in \cite{misselt_jwst_2025}, revealed hundreds of \Hmol\ lines, being the majority of detected lines in the region. 

In this paper, we present the analysis of H$_2$ emission detected with the JWST. In Sect. \ref{Sect2:Data}, we present the observations obtained with the Near-Infrared Spectrograph (NIRSpec) and the Mid-Infrared Instrument - Medium Resolution Spectroscopy (MIRI-MRS), and the data reduction. In Sect. \ref{overview}, we present the detected spectral features and we study the spatial morphology of \Hmol\ emission. In Sect. \ref{extinction}, we use \Hmol\ lines to evaluate the variation of extinction throughout the Horsehead nebula. In Sect. \ref{excitation}, we analyze the excitation of \Hmol\ throughout the PDR to derive physical parameters such as the gas temperature and the ortho-to-para ratio (OPR). In Sect. \ref{discussion}, we discuss the various results of this study by comparing them to template models of the Horsehead nebula.

\section{Observations and Data reduction}
\label{Sect2:Data}

\noindent In this paper, we use MIRI-MRS and NIRSpec from the GTO \#1192 program. The data obtained with MIRI-MRS mode of the Mid-Infrared Instrument \citep[MIRI,][]{wright_mid-infrared_2023} were observed on May 2nd, 2024.
The MRS covers a total wavelength range from $4.9 - 27.9\,\mu $m, separated in four integral field units (IFU) referred to as channels, each divided into three bands. The channels cover slightly different field-of-views (FoV), from 3.2\arcsec$\times$3.7\arcsec (channel 1) up to 6.6\arcsec$\times$7.7\arcsec (channel 4), and have different spatial (from 0.19\arcsec to 0.27\arcsec) and spectral (from $\sim$3700 to $\sim$1500) resolutions \citep{labiano_wavelength_2021}. The observations cover a strip across the Horsehead filaments at the interface between ionized and molecular gas (see Fig.~\ref{fig:footpints_MRS}). We used the 2-point extended source dither pattern, with 26 groups per integration, and one integration per exposure, in FASTR1 readout mode, covering the whole MRS spectral range in three exposures (one per MRS band), for an on-source integration time of 144 s per MRS band. Using the recommended strategy, we took a relatively free-of-emission background observation, with the same integration time per band.
The 2-point dither is not the recommended standard strategy but was rather a compromise given the allocated time in the GTO program. The depth of the background observations is a limiting factor in accurately measuring the continuum \citep[see][for more details]{misselt_jwst_2025}. 
\begin{figure*}
    \centering
    \includegraphics[width=1\linewidth]{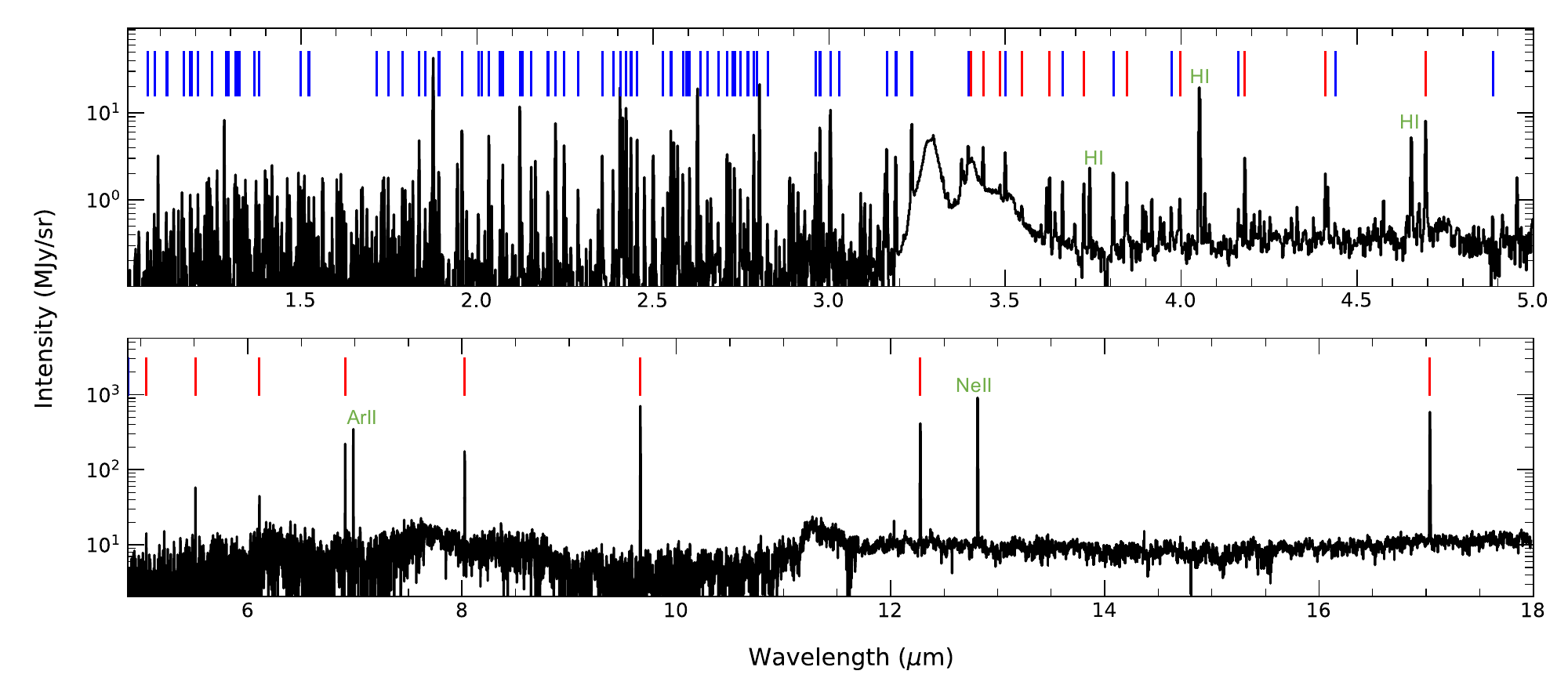}
    \caption{NIRSpec (top) and MIRI-MRS (bottom) spectrum averaged on the first filament, DF1. Red lines (resp. blue lines) correspond to the detected rotational transitions (resp. rovibrational transitions) of \Hmol.  Most of the lines detected in the dissociation front are attributed to \Hmol. The identification of other lines can be found in \cite{misselt_jwst_2025}.}
    \label{fig:spec_h2}
\end{figure*}

The data reduction was done with the JWST Science Calibration Pipeline \citep[version 1.17.1,][]{bushouse_jwst_2023}, with the context 1326 for the Calibration References Data System following the standard procedures \citep[see e.g.,][for detailed examples of MRS data reduction and calibration]{labiano_miri_2016, alvarez-marquez_nuclear_2023}. 
Careful examination of the data showed that the stage 1 corrections \citep{morrison_jwst_2023} could be run with default parameters, and did not leave any significant residuals in the data. 
We switched on the image-to-image background correction in the stage 2 pipeline 
\citep{argyriou_jwst_2023, gasman_jwst_2023, patapis_geometric_2024}. We switched off the master background correction and sky matching steps in stage 3 of the pipeline, before producing the final fully reconstructed science cubes \citep{law_3d_2023} as they introduced artefacts in the data. We then re-aligned the MRS astrometry thanks to the simultaneous imaging data registered to the GAIA DR3 catalog, resulting in less than 0.1" residuals. Finally, the science cubes were shifted to the usual orientation with north up and east to the left, increasing the total FoV in channel 1 to 6.1\arcsec$\times$5.6\arcsec and up to 11.5\arcsec$\times$11.5\arcsec in channel 4.

For NIRSpec, a 3-pointing cycling dither strategy was employed. 4 groups in NRSIRS2 mode (5 frame coadds per downlinked group) were obtained for a total depth of $\sim$875s per pixel in each grating. A background free of emission with dedicated exposures with an identical configuration to a single on-source mosaic pointing was also obtained for NIRSpec. 

The NIRSpec data were processed using the JWST science pipeline version 1.14.0 and the context jwst\_1242.pmap of the Calibration References Data System (CRDS). However, the level-1b rate pipeline was modified slightly to account for a
subtlety in correctly identifying jumps in grouped data with a small number of groups
\citep[see][for more details]{misselt_jwst_2025}. After custom processing, the rate files were re-inserted into the pipeline for level 2 processing. Background subtraction was performed during level 3. 

In the rest of the paper, we use the regions as defined in \cite{misselt_jwst_2025} to analyze \Hmol\ lines. DF1 is defined as the first filament peaking around 2" after the front, and DF2 is defined as the second and third filaments peaking respectively around 8 and 10" after the front 
(see Fig. \ref{fig:footpints_MRS}).

\section{Overview of \Hmol\ emission in the Horsehead nebula}

\label{overview}
\subsection{Detected lines}
In the Horsehead nebula, we detect hundreds of \Hmol\ lines. \cite{misselt_jwst_2025} reveal pure rotational $v=0-0$ ($J<20$) and $v=1-1$ ($J<20$) series with tentative identifications of isolated $v=2-2$, $v=3-3$, and $v=4-4$ lines.  Many ro-vibrational transitions are also detected with transitions in the vibrational $v=1-0$ ($J<16$), $v=2-1$ ($J<19$), $v=2-0$ ($J<8$), $v=3-2$ ($J<14$), and $v=3-1$ ($J<14$) series. Fig. \ref{fig:spec_h2} displays a full NIRSpec and MIRI-MRS spectrum averaged in the first filament, DF1 (see Fig. \ref{fig:footpints_MRS}). This figure shows that \Hmol\ lines correspond to the majority of lines detected in the dissociation front.

\begin{figure*}[htbp]
    \centering
    \includegraphics[width=\linewidth]{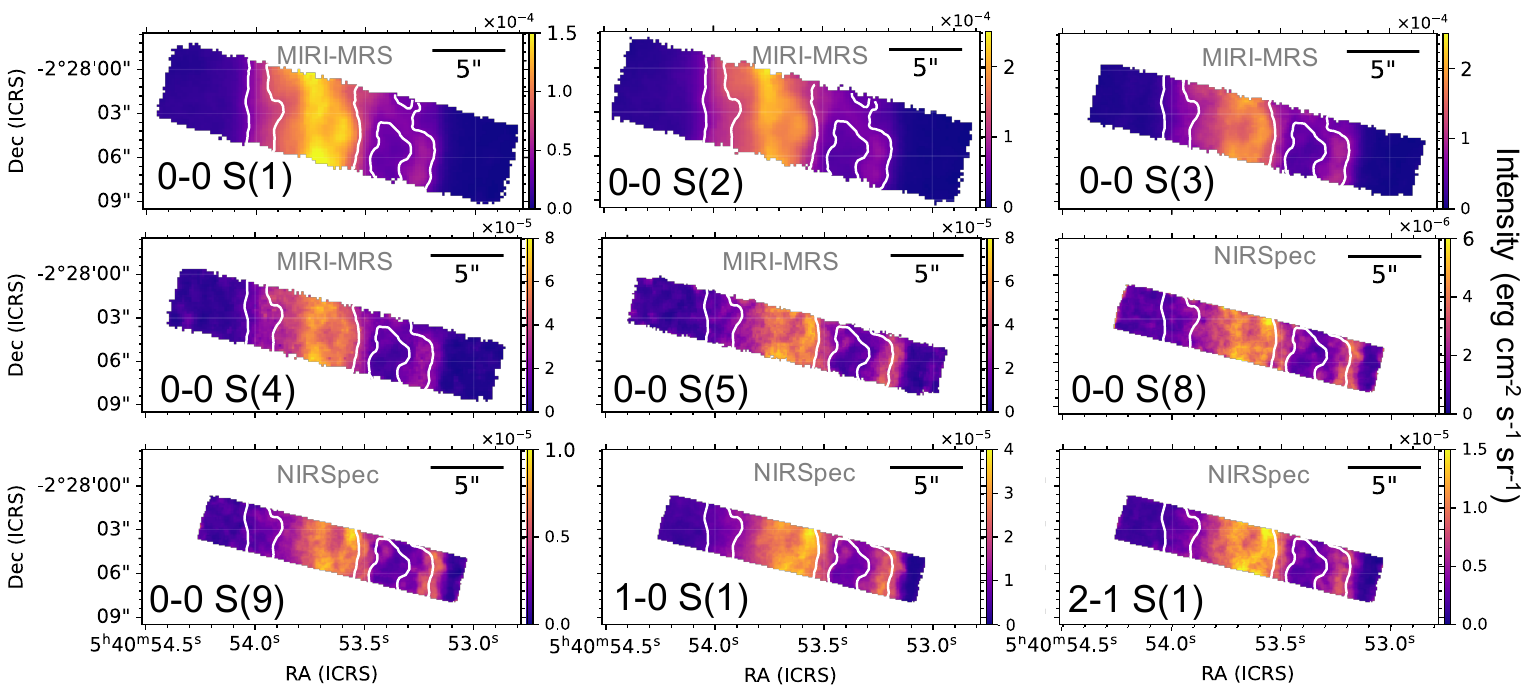}
    \caption{Maps of the brightest \Hmol\ rotational lines emission and the rovibrational lines 1--0 S(1) and 2--1 S(1) emission obtained with MIRI/MRS and NIRSpec across the PDR front. White contours are from the 0--0 S(1) line emission.}
    \label{fig:H2maps}
\end{figure*}

\subsection{Morphology of \Hmol\ emission}

\begin{figure*}
    \centering
    \includegraphics[width=0.82\linewidth]{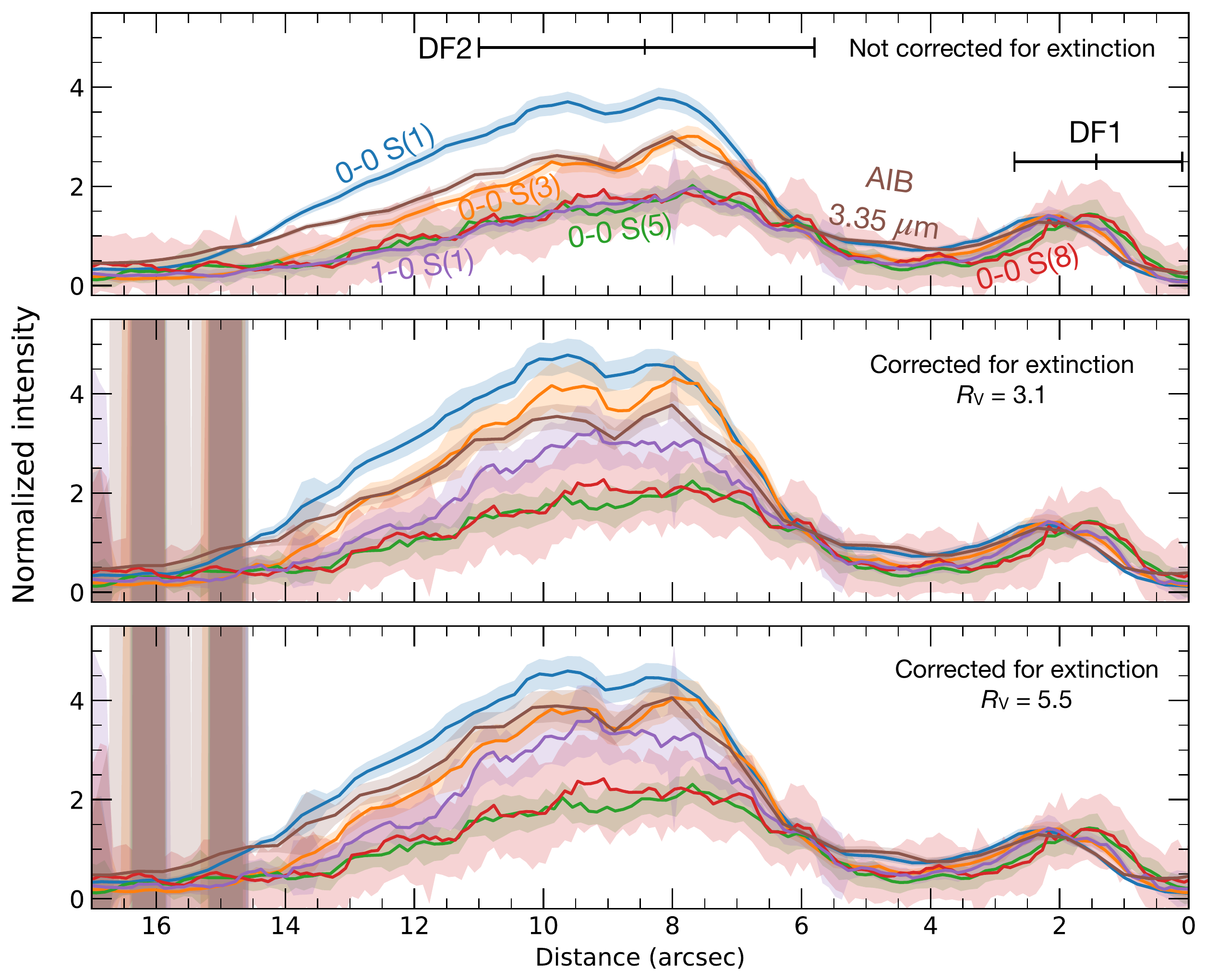}
    \caption{Normalized intensity (around 0.5 and 3" around the first peak) profiles across the front \citep[cut \#3 of][]{abergel_jwst_2024} averaged on 0.5" perpendicular to the line cut. The illuminating star is on the right. (Top) Intensities not corrected for extinction. (Middle and bottom) Intensities corrected for extinction using the attenuation profile derived in Sect. \ref{extinction} using a parametrized extinction curve of \cite{gordon_one_2023} at $R_V = 3.1$ (middle) and $R_V = 5.5$ (bottom). The filled areas are uncertainties. They are particularly important after 15" because the estimation of extinction is very uncertain due to the very low signal-to-noise ratio of the NIR lines used to derive it.}
    \label{fig:H2_cut}
\end{figure*}

The spectro-imaging of the JWST allows us to study the spatial morphology of \Hmol\ emission throughout the Horsehead nebula. To derive the absolute intensities of \Hmol\ lines, we fitted the observed lines with a Gaussian coupled with a linear function to take into account the continuum, and then integrated the Gaussian function over the wavelengths. We chose not to subtract the continuum to minimize the uncertainty of its estimation on the value of the integrated intensity. In this study, only the lines that were nicely reproduced by the fitting procedure were kept (ie, the residuals in the neighborhood of the line after subtraction of the fit were below or similar to the noise in nearby line-free regions).
We were then able to produce maps of the brightest \Hmol\ lines, some displayed in Fig. \ref{fig:H2maps}. This figure shows 9 maps, 7 from pure rotational transitions and two rovibrational transitions. The first five maps are derived from the MIRI-MRS instrument, whereas the last four are derived from the NIRSpec instrument. All maps present spatial structures with widths greater than $\sim1\arcsec$, which is greater than the PSF width at all wavelengths, indicating that these structures are properly resolved for all lines. \Hmol\ emission traces the dissociation front, where atomic hydrogen becomes molecular. Hence, these maps unveil three spatial peaks, corresponding to three distinct dissociation fronts. The two brightest peaks, on the left, 
are separated by $\sim2\arcsec$ and appears on top of a single filament so we will treat them as one in the rest of the paper (corresponding to the DF2 region). These maps reveal a tendency where the more highly excited lines ($J_u > 6$) are peaking closer to the edge of the filaments than the lower excited lines.

Fig. \ref{fig:H2_cut} displays the normalized intensity profile of several \Hmol\ lines throughout the PDR, following the cut \#3 (see Fig. \ref{fig:footpints_MRS}) presented in \cite{abergel_jwst_2024} and averaged on 0.5" perpendicular to the cut. This figure confirms the spatial separation of the \Hmol\ lines, where the FUV-pumped lines (such as 0--0 S(5), 0--0 S(8) and 1--0 S(1)) peak closer to the edge of the PDR than thermalized lines (such as 0--0 S(1), 0--0 S(3)). This spatial separation ($\sim$ 0.5") can be explained by the fact that the first rotational levels of \Hmol\ have lower energy levels, thus they can be excited by collisions at lower temperature. However, higher levels of \Hmol\ are mostly excited by FUV-pumping. Hence, they peak at the position where the UV field is higher, closer to the edge. This figure also unveils that excited \Hmol\ emission peaks occur before nanograin dust emission. Indeed, AIB emission at 3.35 \mum\ (F335M filter of NIRCam) seems correlated with the 0--0 S(1) \Hmol\ line.

\section{Evaluation of the attenuation by the foreground matter}
\label{extinction}
Imaging data presented in \cite{abergel_jwst_2024} have shown that in the JWST wavelength range, the detected emission is affected by attenuation effects due to the foreground matter. To properly analyze \Hmol\ emission, it is thus necessary to correct its emission for extinction, which may vary in the field of view.
 
If the emission is optically thin, which is true for \Hmol\ lines, the intensity of the lines is directly proportional to the column density of the upper level:

\begin{equation}
    I_{ij} = \frac{1}{4 \pi} \frac{hc}{\lambda_{ij}}A_{ij}N_{\rm up}
\end{equation}

\noindent
where $\lambda_{ij}$ is the wavelength of the line, $A_{ij}$ the Einstein coefficient of the line, $N_{\rm up}$ the upper level column density, $h$ the Planck constant and $c$ the light velocity. Thus, the ratio of two lines coming from the same upper level (such as 1--0 S(1) and 1--0 Q(3)) depends only on the wavelengths and the Einstein coefficients of the two lines, and does not depend on physical conditions:

\begin{equation}
    \frac{I_{1}}{I_{2}} = \frac{\lambda_2 A_{1}}{\lambda_1 A_{2}}
    \label{eq:I_ratio}
\end{equation}

\noindent
As the line is attenuated by dust extinction, the observed intensity is written as follows: 

\begin{equation}
    I_{\rm obs} = I \times \exp\left(-\frac{A_\lambda}{A_V} \frac{A_V}{2.5\times \log(e)}\right)
    \label{eq:Iobs}
\end{equation}
\noindent
where $A_V$ is the visual extinction and $A_\lambda/A_V$ the value of the extinction curve at a wavelength $\lambda$. Thus, we can derive the visual extinction from ratios of two \Hmol\ lines coming from the same upper level as:

\begin{equation}
    A_V = \ln\left(\frac{I_{1_{\rm obs}} \lambda_1 A_2}{I_{2_{\rm obs}} \lambda_2 A_1}\right) \frac{2.5\log(e)}{\left(\frac{A_{\lambda_2}}{A_V}-\frac{A_{\lambda_1}}{A_V}\right)}
    \label{eq:AV}
\end{equation}

 To derive the visual extinction, due to the foreground matter, in the different regions of the Horsehead nebula, we used overall 87 \Hmol\ line ratios from the same upper levels. The used transitions have wavelengths going from 1.1 - 4.3 $\upmu$m. There are no lines detected well enough in the MIR with the same upper levels to use them in this analysis. Thus, the attenuation in the NIR biases our estimate of attenuation, and there is no constraint in the MIR. We used this method to derive the visual extinction first in the DF1 and DF2, then throughout the PDR. In order to do that, we used the $R_V$-parametrized extinction curve of \cite{gordon_one_2023} \citep{gordon_fuse_2009,fitzpatrick_analysis_2019,gordon_milky_2021,decleir_spex_2022}  at $R_V = 3.1$, consistent with a diffuse medium extinction curve, and $R_V = 5.5$, consistent with a medium with a depletion of nanograins. In DF1, using the median of the values of $A_V$ derived from the \Hmol\ lines ratios, we find $A_V (R_V = 3.1) = 0.3 \pm 1.3$ and $A_V (R_V = 5.5) = 0.2 \pm 0.5$. Due to the low signal-to-continuum ratio of the lines detected in DF1, the dispersion of the values is quite high, explaining the large uncertainty. Anyway, the value of $A_V$ derived in DF1 is compatible with 0, unveiling the fact that DF1 is attenuated very little. In DF2, we derive $A_V (R_V = 3.1) = 6.1 \pm 1.4$ and $A_V (R_V = 5.5) = 4.5 \pm 1.0$, showing that DF2 is a lot more attenuated than DF1. The values derived are in agreement with the estimate of $A_V$ of \cite{misselt_jwst_2025} from HI recombination lines. They find $A_V = 1.51 \pm 0.93$ in DF1 and $A_V = 7.02 \pm 2.60$ in DF2. 

\begin{figure}
    \centering
    \includegraphics[width=\linewidth]{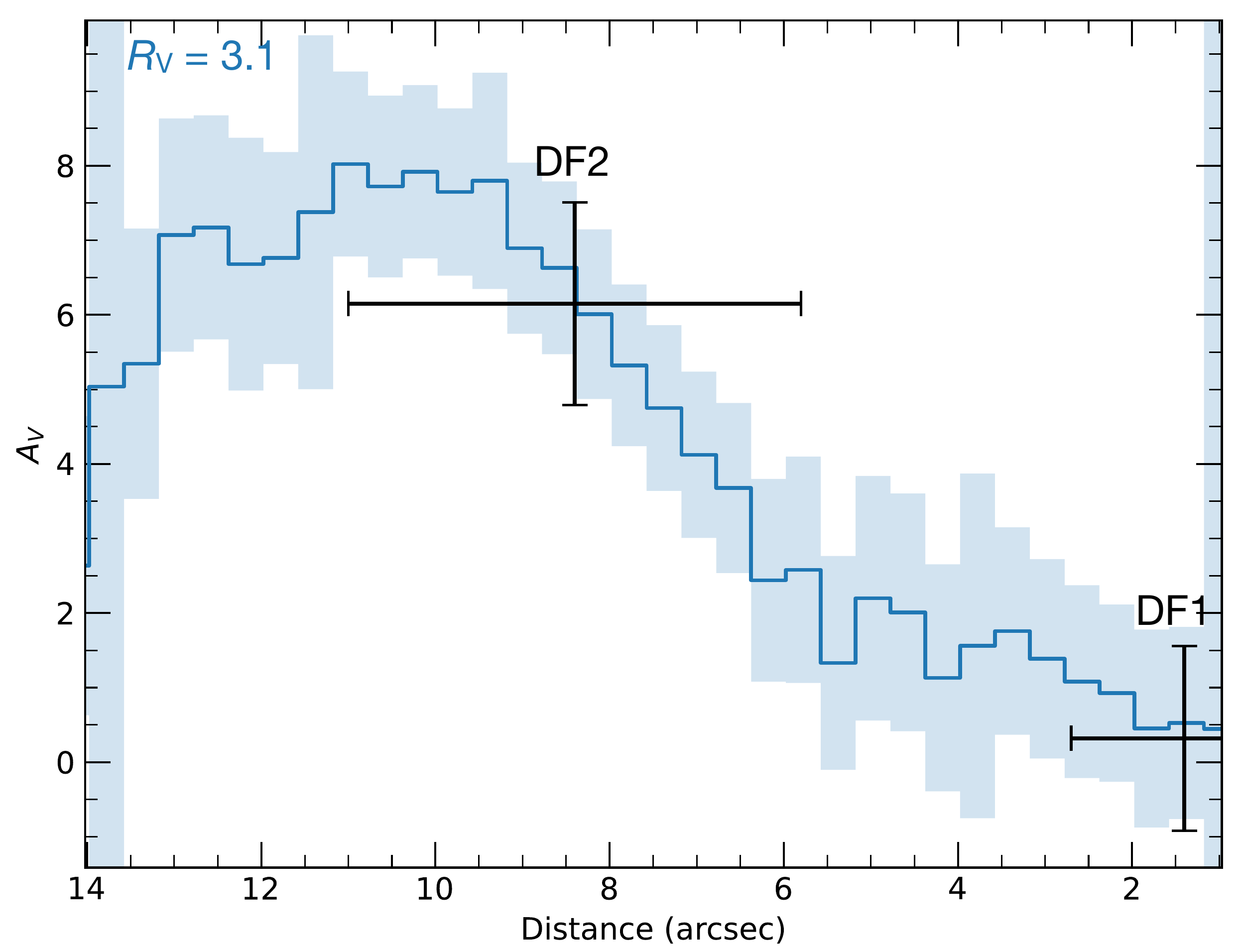}
    \caption{Profile of $A_V$, the attenuation by the foreground matter, across the PDR front derived from \Hmol\ line ratios. The signal was averaged on the width of the maps and over four columns of pixels to compute \Hmol\ maps with sufficient SNR. Before 1" and above 14" from the front, \Hmol\ emission is too faint to derive $A_V$. $A_V$ is increasing from the edge of the PDR to the second and third \Hmol\ filament.}
    \label{fig:Av_cut}
\end{figure}

In order to estimate the foreground attenuation $A_V$ throughout the PDR, we
use maps from \Hmol\ rovibrational lines. However, as they are quite weak, we had to degrade the spatial resolution of the NIRSpec data cube. Thus, we averaged the signal following the width of the map and also over four columns of pixels. In every column of pixels, we use as many detected line ratios as possible to derive an $A_V$ value. Fig. \ref{fig:Av_cut} displays the $A_V$ profile, using an extinction curve estimated at $R_V = 3.1$ throughout the PDR, following cut \#3 of \cite{abergel_jwst_2024}. The value of $A_V$ increases from the first \Hmol\ and dust front to the second and third ones. A similar profile is obtained when using $R_V = 5.5$ with $A_V$ value being 1.4 times lower. Below a distance of 1" and above 14" from the first front, the intensity of \Hmol\ drops drastically, making it impossible to estimate the extinction in those regions. This $A_V$ profile is also consistent with the profile derived in \cite{misselt_jwst_2025} from HI recombination lines coming from the same wavelength range (see their Fig. 6). 

These results are consistent with the schematic view of the Horsehead geometry displayed in Fig. 13 of \cite{abergel_jwst_2024}, with the PDR illuminated from behind. Indeed, the higher extinction in DF2 than in DF1 seems to highlight the fact that DF2 is located further away from the observer than DF1. It also confirms that the \Hmol\ emitting region is very small compared to the depth of the material that is attenuating the signal. Indeed, from the \Hmol\ column density we derive in Sect. \ref{excitation}, the emitting region of \Hmol\ is expected to peak at $A_V < 0.2$.

\begin{figure}
    \centering
    \includegraphics[width=\linewidth]{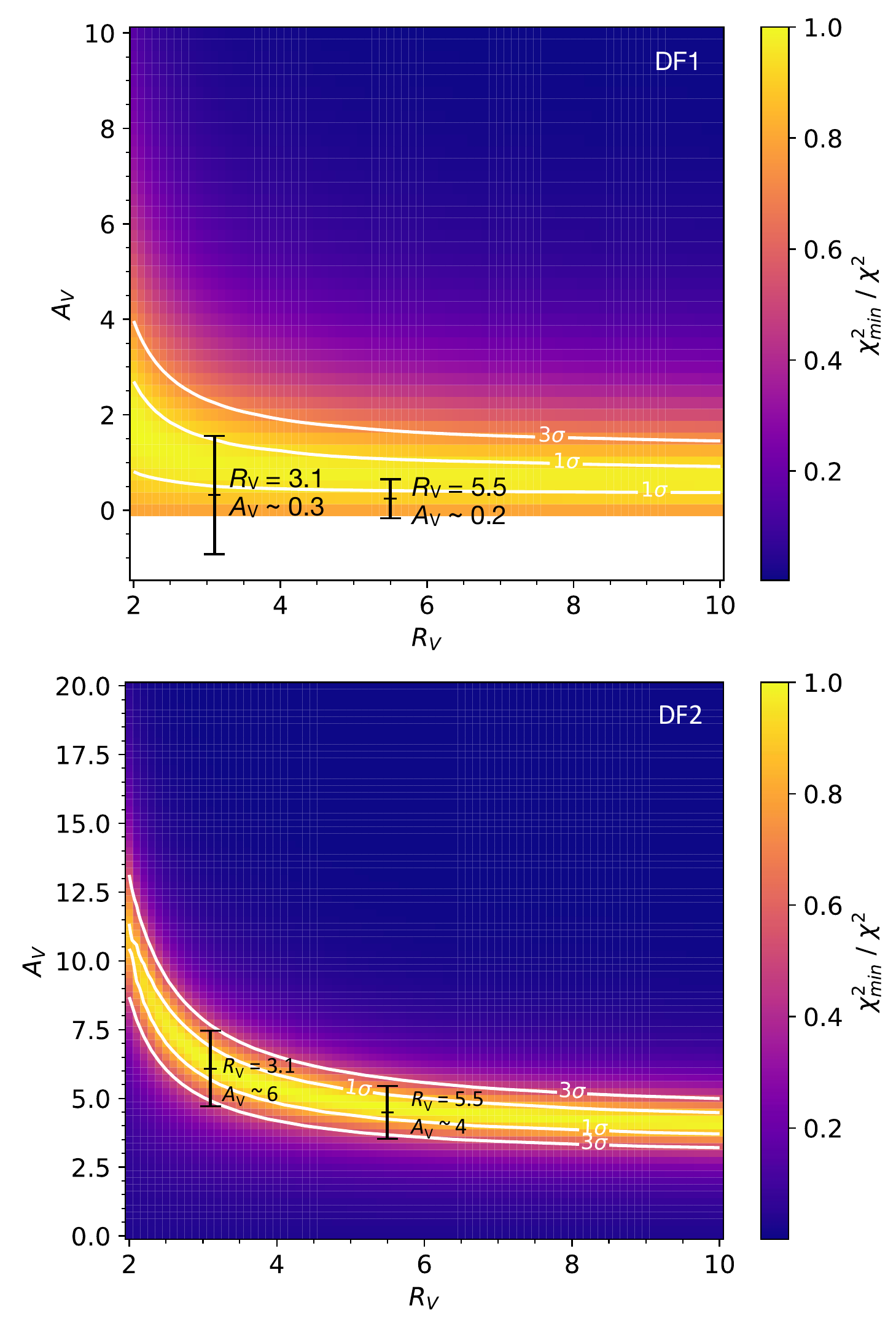}
    \caption{$\chi^2_{\rm min}$/$\chi^2$ maps of the difference between theoretical \Hmol\ ratios and observed \Hmol\ ratios corrected for extinction in DF1 and DF2 as a function of $A_V$ and $R_V$. The white lines are the contours of valid values at 1$\sigma$ and 3$\sigma$. $A_V$ and $R_V$ are degenerate, so we can not discriminate the extinction curve in the Horsehead nebula.}
    \label{fig:chi2_Av_Rv}
\end{figure}

In order to discriminate between different extinction curves (e.g., $R_V = 3.1$ or $R_V = 5.5$) in the Horsehead nebula, we used a $\chi^2$ method. We compared the theoretical \Hmol\ ratio, given in Eq. (\ref{eq:I_ratio}), to the observed ratio when corrected for extinction using extinction curves with parameters varying from $A_V = 0 - 20$ and $R_V : 2-10$. We calculated $\chi^2$ as:

\begin{equation}
    \chi^2 = \frac{\left(\frac{I_{1_{\rm obs}}}{I_{2_{\rm obs}}}-\frac{I_1}{I_2}\right)^2}{\sigma^2}
\end{equation}
where $\frac{I_{1_{\rm obs}}}{I_{2_{\rm obs}}}$ is the ratio corrected for extinction, $\frac{I_1}{I_2}$ the theoretical \Hmol\ ratio, and $\sigma$ the uncertainty in the observed ratio\footnote{$\sigma=\frac{I_{1_{\rm obs}}}{I_{2_{\rm obs}}}\sqrt{\left(\frac{\sigma_1}{I_{1_{\rm obs}}}\right)^2 + \left(\frac{\sigma_2}{I_{2_{\rm obs}}}\right)^2}$}. Fig. \ref{fig:chi2_Av_Rv} shows that $R_V$ and $A_V$ are highly degenerate, making it impossible to discriminate the extinction curve in the Horsehead nebula. This is mainly because we only use \Hmol\ lines in the NIR, which does not allow lifting of the degeneracy.

The middle and bottom panels of Fig. \ref{fig:H2_cut} show the normalized intensity profile when correcting for extinction using our previous estimate of $A_V$ using extinction curves from \cite{gordon_one_2023} evaluated at $R_V = 3.1$ and $R_V = 5.5$. These panels reveal that the attenuation we derive cannot account for the variation of \Hmol\ lines intensity ratios from DF1 to DF2. For thermalized lines (0--0 S(1) and S(3)), the difference can then be explained by the variation of temperature (see Sect. \ref{temperature}). However, we expect the FUV-pumped lines (0--0 S(5), 0--0 S(8) and 1--0 S(1)) to have higher intensity where the column density is higher (hence DF2). The figure shows that if the intensity of the 1--0 S(1) increases by a factor of 2.5 between the first and second filament, the intensity of the 0--0 S(5) and 0--0 S(8) lines only increases by a factor below 2. Our correction of extinction does not increase the intensity of these lines in DF2. In addition, dust profiles at similar wavelengths (5.5 \mum\ and 7 \mum) display a second filament 3 times higher than the first \citep[see Fig. 7 of][]{abergel_jwst_2024}. Hence, these characteristics cannot be explained by extinction effects. This seems to show that the pure rotational levels, even excited, are more sensitive to the temperature variation than rovibrational levels. These results may indicate that excitation by collision in the first filament is not negligible for highly excited rotational levels.

\section{\Hmol\ excitation and physical state of the filaments}
\label{excitation}
\subsection{Estimate of the gas temperature}
\label{temperature}

\begin{figure}
    \centering
    \includegraphics[width=\linewidth]{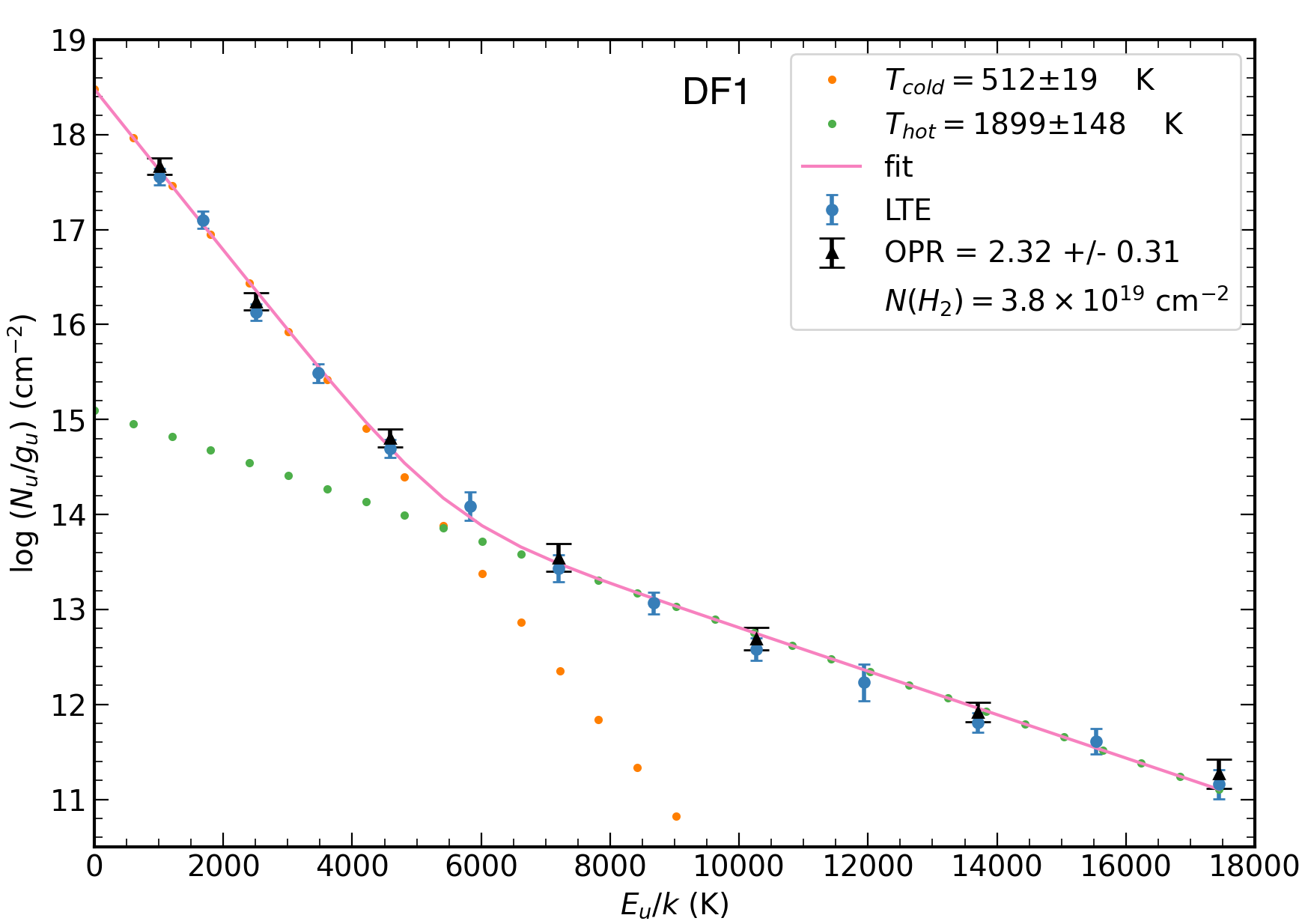}

    \includegraphics[width=\linewidth]{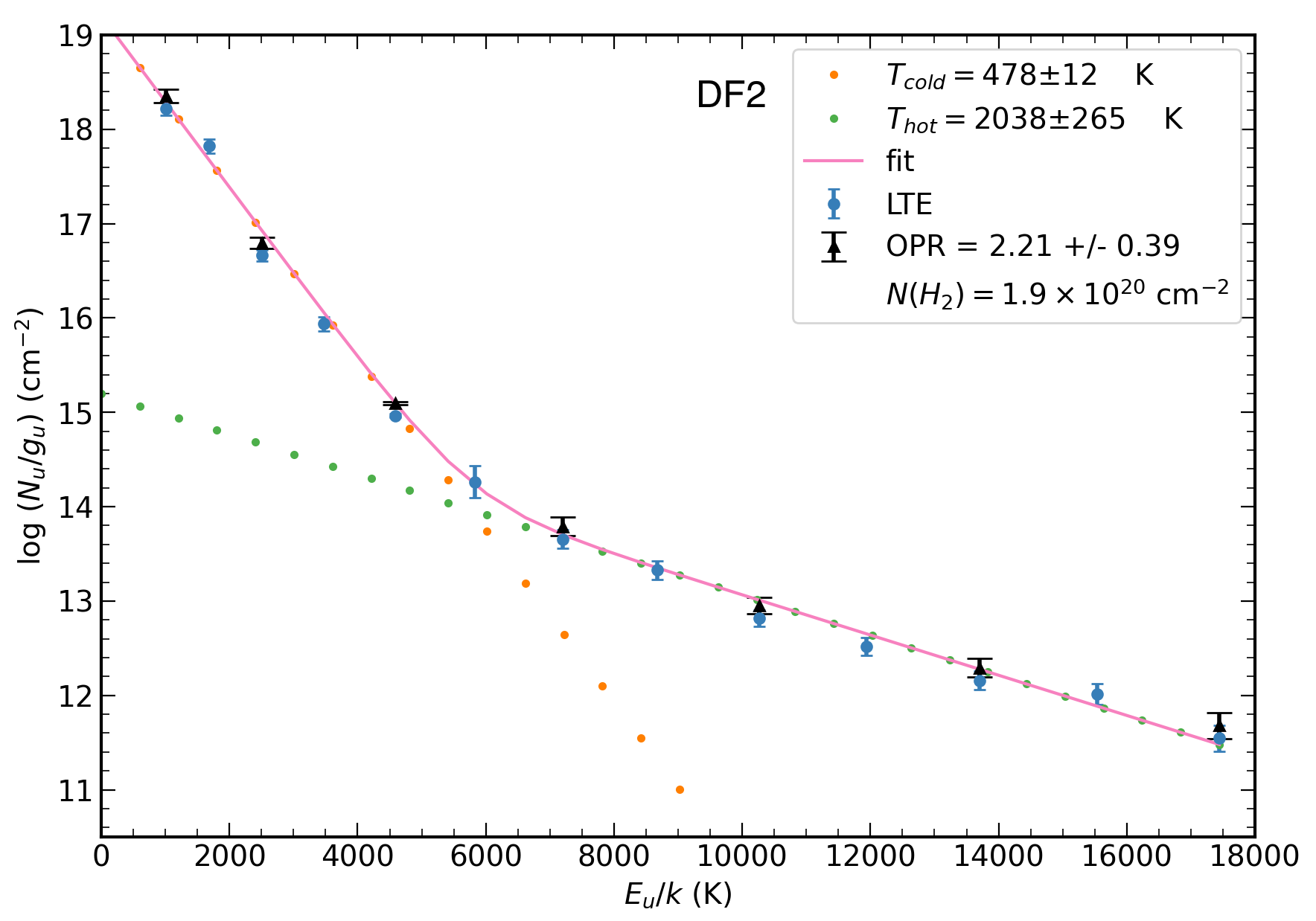}
    \caption{Excitation diagram (top) in the DF1 (bottom) in the DF2. Full excitation diagrams are presented in Appendix \ref{app:excitation_diagram}.}
    \label{fig:diag_rot}
\end{figure}

In this section, we discuss the excitation of \Hmol\ throughout the PDR. First, we derive the gas temperature in the different dissociation fronts using the apertures defined in \cite{misselt_jwst_2025}, DF1 and DF2 (see Fig. \ref{fig:footpints_MRS}). Fig. \ref{fig:diag_rot} displays the excitation diagram of the $v=0$ levels of \Hmol\ in the DF1 and DF2. These excitation diagrams were done using the \texttt{PhotoDissociation Region Toolbox Python module}\footnote{\url{https://github.com/mpound/pdrtpy}} \cite[pdrtpy,][]{kaufman_si_2006,pound_photo_2008,pound_pdrt_2011,pound_photodissociation_2023}. For DF2, the line intensities used to plot the excitation diagram were corrected for extinction using the $R_V$-parametrized extinction curve of \cite{gordon_one_2023} estimated at $R_V = 3.1$ and considering $A_V = 6.1$ as derived in Sect. \ref{extinction}. In both regions, the excitation diagrams unveil two temperatures. This means that the excitation of \Hmol\ follows two Boltzmann distributions as described by:

\begin{equation}
\ln\left(\frac{N_{\rm up}}{g_{\rm up}}\right)= \ln\left(\frac{N}{Q(T_{\rm ex})}\right)-\frac{E_{\rm up}}{k_{\rm B} T_{\rm ex}},
\label{eq:excit_diagram}
\end{equation}

The slope break occurs at an energy of 6000 K, precisely where the first levels of $v = 1$ are located. As the first levels of \Hmol\ ($J<6$) are maintained in thermal equilibrium via collisions, the cold excitation temperature corresponds to the gas temperature. The excitation diagrams presented in Fig.\,\ref{fig:diag_rot} show that in DF1, the observed gas temperature is about $T_{\rm obs} = 512 \pm 19$~K. In the DF2, the observed gas temperature is slightly lower at $T_{\rm obs} = 478 \pm 12$ K. The difference in gas temperature between the two apertures is likely due to the fact that the front of DF2 is mixed with the decrease of DF1. Thus, we do not have access to the high-temperature part of the front, similarly to DF1. Using Eq. (\ref{eq:excit_diagram}), we can also derive the total column density of \Hmol\ from these excitation diagrams. The column density in DF1 is $N({\rm H}_2) = (3.8 \pm 0.8) \times 10^{19}$ cm$^{-2}$ and in DF2, $N({\rm H}_2) = (1.9 \pm 0.4) \times 10^{20}$ cm$^{-2}$. Hence, we revealed that the column density of \Hmol\ is 5 times higher in DF2 than in DF1. A summary of the derived parameters is presented in Table \ref{tab:results}.

\begin{figure}
    \centering
    \includegraphics[width=\linewidth]{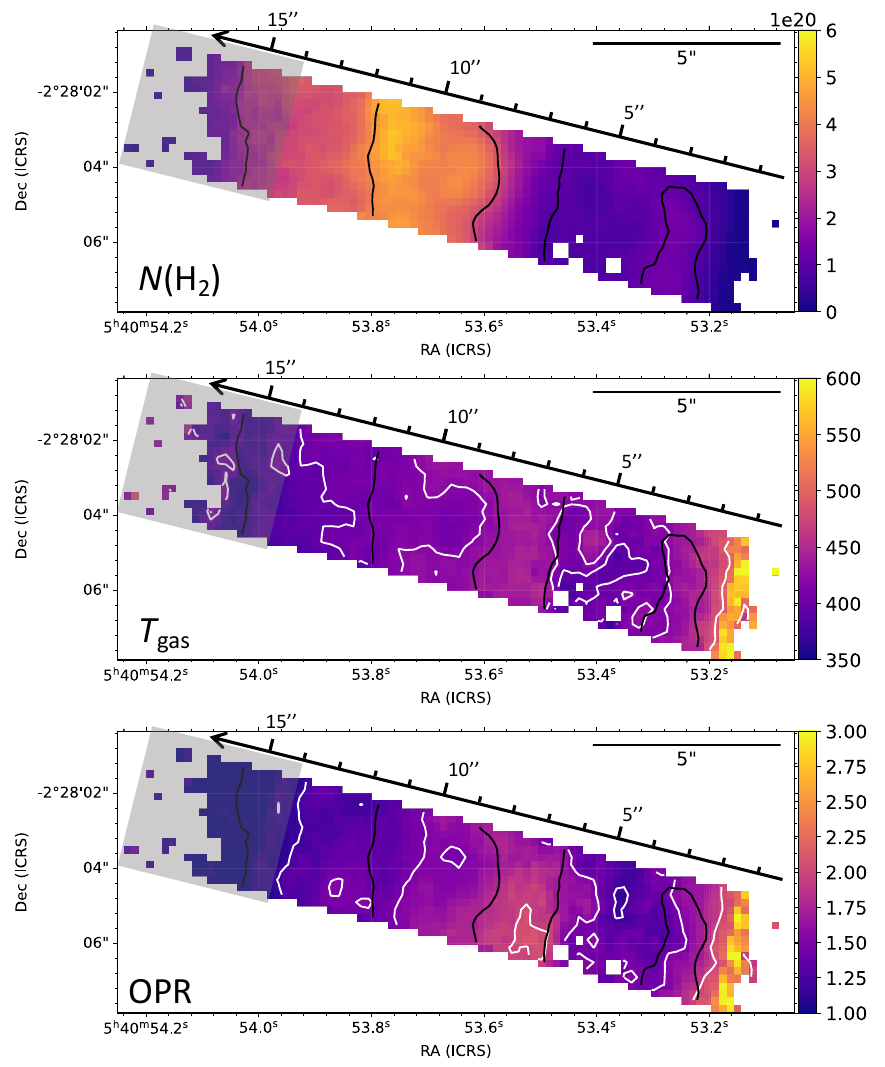}
    \caption{(Top panel) Column density,  (Middle panel) gas temperature, (Bottom panel) Ortho-to-para ratio map corrected for extinction. Black contours are \Hmol\ 0--0 S(1) line corrected for extinction (levels : 4.5 $\times$ 10$^{-5}$ , 1.5 $\times$ 10$^{-4}$ erg cm$^{-2}$ s$^{-1}$ sr$^{-1}$). The white contours levels are 400, 420, 500 K for $T_{\rm gas}$ and 1.2, 1.5, 2 for the OPR. The gray boxes overlayed on the map correspond to the region where the extinction is badly constrained.}
    \label{fig:map_temp_coldens}
\end{figure}

We also derived the gas temperature and column density across the entire MIRI-MRS map using the pdrtpy toolbox. We cannot fit two temperature components, as we did in averaged regions (DF1 and DF2) because we cannot produce enough maps of \Hmol\ lines with sufficient signal-to-noise ratio. Hence, we fitted the cold component of the excitation diagram in each pixel of MIRI-MRS, using the maps of the first four lines of \Hmol\ detected with JWST (0--0 S(1) - 0--0 S(4)), which are thermalized. The column density and gas temperature maps are displayed in Fig. \ref{fig:map_temp_coldens}. In the top panel of this figure, we find that the column density is the highest around the second and third filament($N(\Hmol) \sim 5\times 10^{20}$ cm$^{-2}$, with a mean pixel relative error of $\frac{\sigma_{N(\text{H}_2)}}{N(\text{H}_2)} \sim 10\%$), agreeing with the value derived from the excitation diagrams in Fig. \ref{fig:diag_rot}. The middle panel shows that the gas temperature does not vary much throughout the PDR (from 600 K to 380 K, with a mean pixel relative error of $\frac{\sigma_{T_{\rm obs}}}{T_{\rm obs}} \sim 2\%$). This is in contradiction with the expected attenuation of the UV field, and thus the expected temperature decrease (down to $\sim 100$ K), inside each filament.

\begin{figure}
    \centering
    \includegraphics[width=\linewidth]{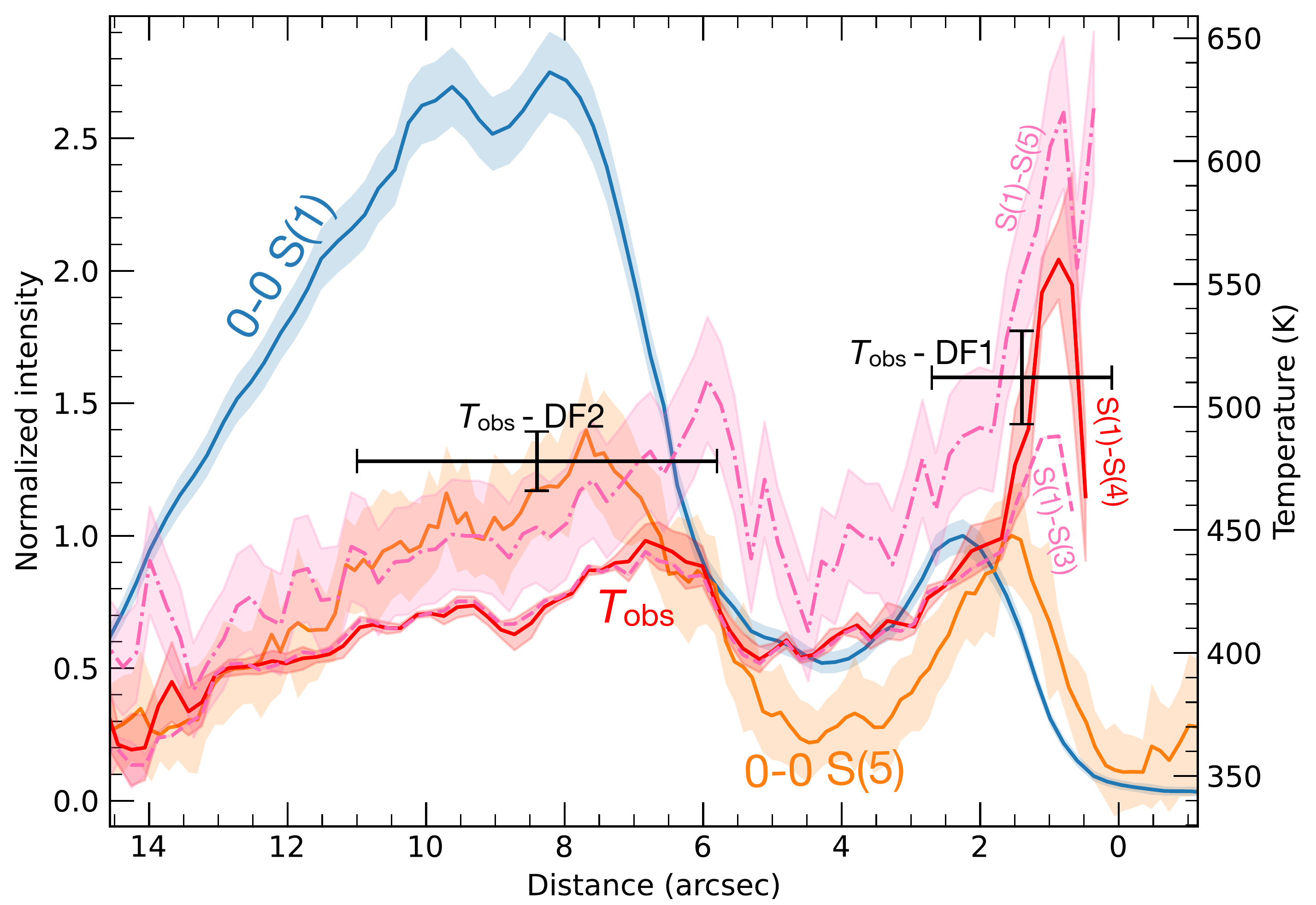}

    \caption{Gas temperature, derived from the first four observed lines of \Hmol\ (S(1) - S(4)) profile compared to \Hmol\ 0--0 S(1) and 0--0 S(5) line emission profile. The dashed pink line (resp. dashed-dotted pink line) corresponds to the gas temperature derived from the first three observed lines (S(1) - S(3)) (resp. the first five observed lines (S(1) - S(5)). We observe a slight decrease in temperature in the \Hmol\ filament and a similar temperature at each dissociation front. }
    \label{fig:cut_temp}
\end{figure}

Fig. \ref{fig:cut_temp} shows the observed temperature profile following cut \#3 of \cite{abergel_jwst_2024}. This figure reveals that the temperature decreases after the peak of each filament. The temperature derived from the first four lines of H$_2$ reveals a decrease from 550 K  to 400 K in the first filament and from 450 K to 400 K in the second and third filaments. The temperature at the peak of DF1 and DF2 is similar and around 440 K. The dashed pink line represents the temperature derived from the first three lines (S(1)-S(3)). The estimated temperatures taking into account 4 or 3 \Hmol\ lines are similar at the peak of DF1 and in DF2, but there is a difference in front of DF1. Indeed, at the front of the PDR, the temperature derived from the first four lines is around $T_{\rm obs} \sim 550$ K against $T_{\rm obs} \sim 480$ K with the first three lines. This reveals the impact of UV-pumping in the excitation of the $J=6$ level, which might not be thermalized at the front.

\subsection{Spatial variations of the ortho-to-para ratio}

With the pdrtpy module, we also fitted the OPR. The excitation diagram in Fig. \ref{fig:diag_rot} shows that the OPR is not at the expected value at equilibrium, which is 3 for temperatures above 200~K \citep{sternberg_ratio_1999}. The value in both DF1 and DF2 is close to OPR $\sim$ 2. Fig.~\ref{fig:map_temp_coldens} displays a map of the OPR in the Horsehead nebula. Overall, the value of the OPR is lower than 3 everywhere in the Horsehead. The variation displayed in the map shows that the OPR reaches its highest value (OPR $\sim$ 2-2.5) at the edge of each dissociation front, where the temperature increases. Deeper into the PDR, the OPR decreases to a smaller value around OPR $\sim$  1.3-1.5. The variations are significant as the mean (resp. median) error per pixel is around $\sigma_{\rm OPR}(\text{mean}) \sim 0.3$ (resp. $\sigma_{\rm OPR}(\text{median}) \sim 0.06$). These OPR values are in contradiction with the high temperatures detected, where the equilibrium should be easily reached. This reveals the presence of out-of-equilibrium mechanisms such as the ortho-para conversion, favoring the formation of para \citep{bron_efficient_2016} or the advection and mixing of colder \Hmol\ in the warm region \citep{gorti_photoevaporation_2002,storzer_nonequilibrium_1998}. The latter explanation is in agreement with the photoevaporating flow observed in the imaging data, revealing important dynamical effects \citep{abergel_jwst_2024}. 

Interestingly, we also observe an inversion of the 0--0 S(1) and 0--0 S(2) peaks (see top panel of Fig. \ref{fig:cut_ortho_para}). These results confirm the great efficiency of ortho-para conversion, which leads to small OPR inside the PDR, both with gas phase reactive collisions (when the gas temperature decreases) and dust surface conversion (when the dust temperature decreases) \citep{bron_efficient_2016}. The spatial shift between ortho and para levels is not seen for more excited \Hmol\ rotational levels as they originate from warmer regions where the OPR is close to 3. Besides, the OPR derived from excited rotational levels ($J_{\rm up}=5-7$) is higher and closer to 3 than the lowest levels ($J_{\rm up}=3-5$) (this is somewhat visible on the excitation diagrams in Fig. \ref{fig:diag_rot} with the high $J$ levels better aligned than the low $J$ levels).

In addition, a different OPR is observed for the rovibrational transitions. Indeed, as observed in the excitation diagrams, the OPR for the rovibrational levels is lower than for the rotational levels (see Fig. \ref{fig:diag_rot_tot}). We observe  $\mathrm{OPR}_{\rm vib} \sim \sqrt{\mathrm{OPR}_{\rm rot}}$ throughout the map (see Fig. \ref{fig:opr_rovib}). This can be explained by a preferential self-shielding of ortho levels compared to para levels. This process favors the pumping of para levels and thus the para vibrational transitions \citep{sternberg_ratio_1999}. That is why we also observe a spatial shift between ortho and para levels in the rovibrational transitions (see Fig. \ref{fig:map_10S1_10O2}, Fig. \ref{fig:spec_region} and bottom panel of Fig. \ref{fig:cut_ortho_para}).

\begin{table*}
    \centering
    
    \caption{Parameters derived from \Hmol\ lines in the Horsehead nebula.}
    \begin{tabular}{c|c|c|c|c|c}
\multirow{2}{*}{Region} & $A_V$  & $A_V$  & Observed  & Observed column  & Observed    \\
 & (from \Hmol\ lines) & (from HI lines) & temperature $T_{\rm obs}$ (K) &   density $N_{\rm obs}$ (cm$^{-2}$) &  OPR   \\
\hline
\hline
DF1  & 0.3 $\pm$ 1.3 & 1.51 $\pm$ 0.93 & 512 $\pm$ 19 & (3.8 $\pm$ 0.8) $\times$ 10$^{19}$ & 2.3\\
\hline
DF2 & 6.1 $\pm$ 1.4 & 7.02 $\pm$ 2.60 & 478 $\pm$ 12 & (1.9 $\pm$ 0.4) $\times$ 10$^{20}$ & 2.2
    \end{tabular}
    \label{tab:results}
\end{table*}

\section{Discussion}

\label{discussion}

In the previous section, we derived the temperature profile throughout the PDR. In this section, we discuss the resulting estimated thermal pressure and present brief comparisons with template stationary 1D PDR models to better understand the observed profile.

\subsection{Geometry of the Horsehead nebula}
\label{geometry}

\begin{figure}
    \centering
    \includegraphics[width=\linewidth]{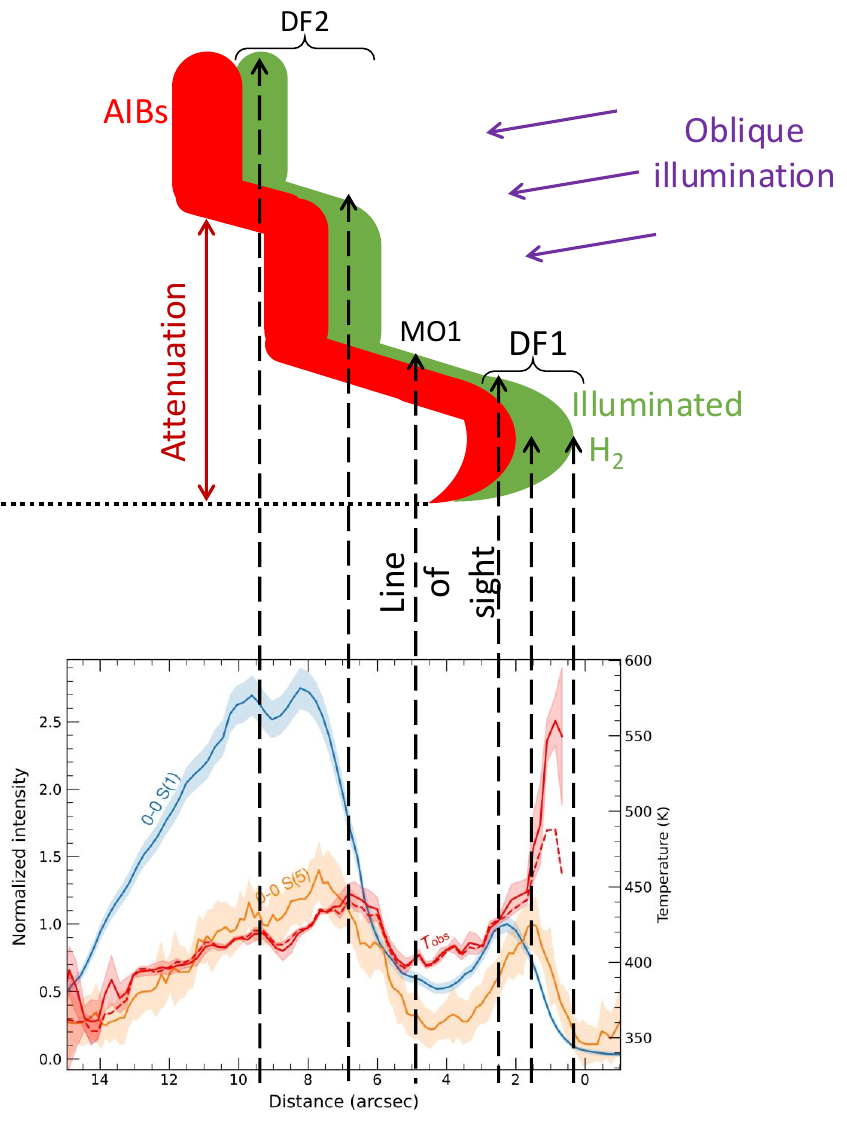}
    \caption{Comparison between the schematic view of the geometry of the Horsehead nebula and the observed profile of \Hmol\ and temperature. Figure adapted from \cite{abergel_jwst_2024}.}
    \label{fig:geometry_HH}
\end{figure}

The Horsehead is slightly illuminated on its backside as shown in Fig.~\ref{fig:geometry_HH} and thus is not entirely edge-on. This implies that no matter the distance from the edge of the PDR (here taken as the origin), there is always illuminated material on the backside. This geometry explains the low variation in temperature inside the PDR, which is expected to decrease quite rapidly between the peaks of 0--0 S(5) and 0--0 S(1). Indeed, as the \Hmol\ first rotational levels are very sensitive to the temperature, their emission is dominated by the illuminated matter. The observed temperature then reflects the mean gas temperature in the \Hmol\ emitting region. Hence, it is highly probable that the real gas temperature at and after the 0--0 S(1) peak is actually lower.

\subsection{Estimate of thermal pressure}

Due to the complex geometry of the Horsehead, it is very difficult to estimate the local thermal pressure inside the PDR. As the observed temperature is dominated by a thin layer of illuminated \Hmol, only the observed temperature at the very edge of the PDR ($d < 1.5$", $A_V \sim 0.1$), where all the gas is illuminated, can be attributed to the local gas temperature.

To estimate the thermal pressure at the edge of the PDR, we use the measured \Hmol\ column density to derive an estimate of the local gas density. At 1.5" from the edge, the column density of \Hmol\ is around $N$(\Hmol) $\sim 6 \times 10^{18}$ cm$^{-2}$. We consider that at the edge, the size of \Hmol\ emission in the line of sight is similar to the distance from the edge toward the star, ie, around 1.5". With this length, we can estimate \Hmol\ density $n(\Hmol) \sim 6 \times 10^3$ \cc. At this position, the temperature is around 480 K. Thus, we derive \Hmol\ thermal pressure around $P(\Hmol) = n(\Hmol) T_{\rm gas} \sim 3 \times 10^6$ K \cc. If we consider that all the hydrogen is molecular, then 
$n = n(\text{H}_2)$ and we can estimate a lower limit of the thermal pressure $P_{{\rm th}_{low}} = nT_{\rm gas} = n(\Hmol)T_{\rm gas} \sim 3 \times 10^6$ K \cc. However, at this distance from the edge, we expect that a lot of the hydrogen is still atomic or ionized. Hence, if we consider that the emission originates from a region before the H/\Hmol\ transition\footnote{Here, we define the H/\Hmol\ transition as the position where $x$(H)=$x$(\Hmol).} then $x$(H) $\geq$ $x$(\Hmol) and $n = n(\text{H}) + n(\Hmol) \geq 2 n(\Hmol)$. Finally, we find that the thermal pressure is at least $P_{\rm th} \geq 2n(\Hmol)T_{\rm gas} \sim 6 \times 10^6$ K \cc.

This value is slightly higher than previous estimates, which are around $P_{\rm gas} \sim (2-4) \times 10^6$ K \cc \citep[e.g.,][]{habart_density_2005,hernandez-vera_extremely_2023}.
This can be explained by the fact that the measure of the pressure was not made at the same position in these studies. In fact, \cite{hernandez-vera_extremely_2023} derive the thermal pressure using CO and HCO$^+$ ALMA data at $\delta x = 15"$ from the edge. At this position, they derive, from HCO$^+$ emission, a higher gas density ($n_{{\rm H}_{\rm CO}} = (3.9 - 6.6) \times 10^4$ \cc) than we estimate at the edge ($n_{{\rm H}_{\rm H_2}} \sim 10^4$ \cc). However, from the CO emission, they derive a very low gas temperature $T_{{\rm gas}_{\rm CO}} = 40-60$ K. The cold gas they are probing, deeper into the PDR, could have a lower thermal pressure than the hot gas at the edge.

In this study, they also derive a thermal pressure by comparing the position of the H/\Hmol\ and the C/CO transitions with PDR models. They use as constrains $d_{\rm H/H_2} < 650$ au and $d_{\rm C/CO} \sim 1200$ au (see their Fig. 8). From this method, they derive higher thermal pressures, $P_{\rm gas} = (3.7-9.2) \times 10^6$ K \cc\, which are more compatible with our estimate. In addition, JWST observations reveal that the atomic layer is even smaller than expected, such as $d_{\rm H/H_2} < 100$ au \citep{abergel_jwst_2024}, pointing to high thermal pressures.

The estimate of the thermal pressure from \Hmol\ lines $P_{{\rm th}} \geq 6 \times 10^6$ K \cc\ is higher than the pressure predicted in the adjacent \Hii region IC 434 $P_{\rm th, \Hii} \approx n_e T_e = (2.4 - 8.0) \times 10^5$ \cc \citep{bally_kinematics_2018}. This overpressure observed between the edge of the PDR and the \Hii region could explain the photoevaporating flow observed in the imaging data \citep{abergel_jwst_2024}. We discuss this in Sect.~\ref{subsec:dynamics}.

\subsection{Comparison with 1D stationary PDR models}

In order to compare the observations with stationary 1D PDR models, we use the online grid of isobaric models of the Meudon PDR code\footnote{Grid of isobaric models from August 2024: \url{https://app.ism.obspm.fr/ismdb/}} \citep[][version 7]{le_petit_model_2006}. The code simulates the thermal and chemical structure of the gas in a self-consistent manner, considering a 1D geometry and a stationary state in a plane-parallel irradiated gas and dust layer. The incident UV radiation field is that of an O5 star with an intensity of $G_0 = 100$. The code includes progressive attenuation of the UV field as a result of grain and gas extinction. The extinction curve used is the mean galactic extinction curve with the parameterization of \cite{fitzpatrick_analysis_1988}.

\begin{figure}
    \centering
    \includegraphics[width=\linewidth]{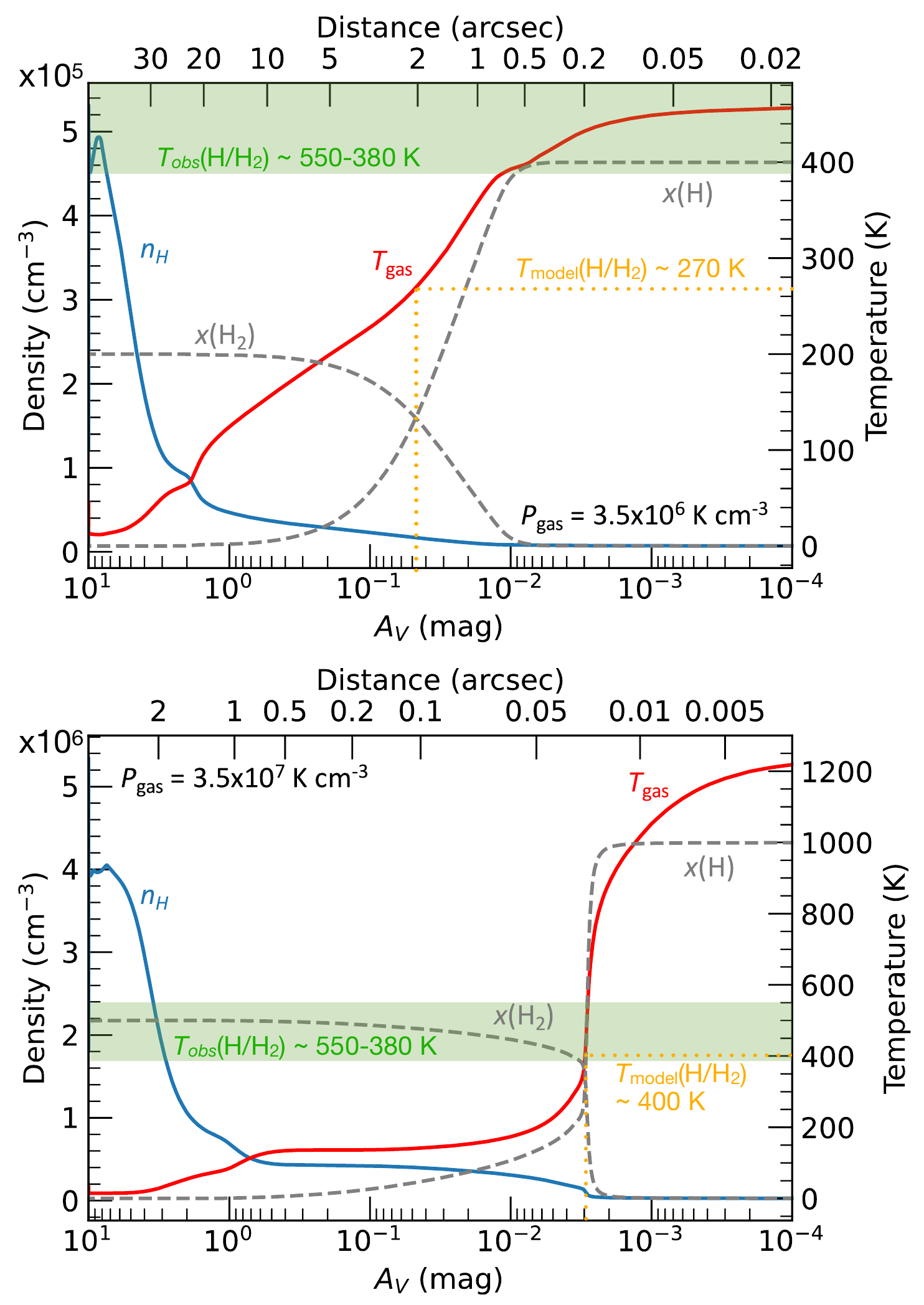}
    \caption{Density (blue) and temperature (red) profile as a function of $A_V$, the visual extinction within the PDR\protect\footnotemark and distance (pc)\protect\footnotemark  in 1D stationary PDR models with $G_0 = 100$ and (Top) $P_{\rm gas} = 3.5\times10^6$ K \cc. (Bottom) $P_{\rm gas} = 3.5\times10^7$ K \cc. The gray dashed lines correspond to the abundance of H and \Hmol.}
    \label{fig:nH_T}
\end{figure}

Figure \ref{fig:nH_T} displays the density and temperature profiles in PDR models at different thermal pressures. This figure shows that these template 1D stationary PDR models, no matter the thermal pressure, cannot reproduce the temperature profile observed in the Horsehead nebula. 
Indeed, in models, the gas temperature always decreases to values way below 400 K inside the PDR. This difficulty in reproducing the temperature profile is in agreement with the argument presented in Sect. \ref{geometry}, that the temperature derived with \Hmol\ lines is not a good tracer of the gas temperature deep inside the PDR, but only reflects the value at the H/\Hmol\ transition.

\begin{figure}[!h]
    \centering
    \includegraphics[width=\linewidth]{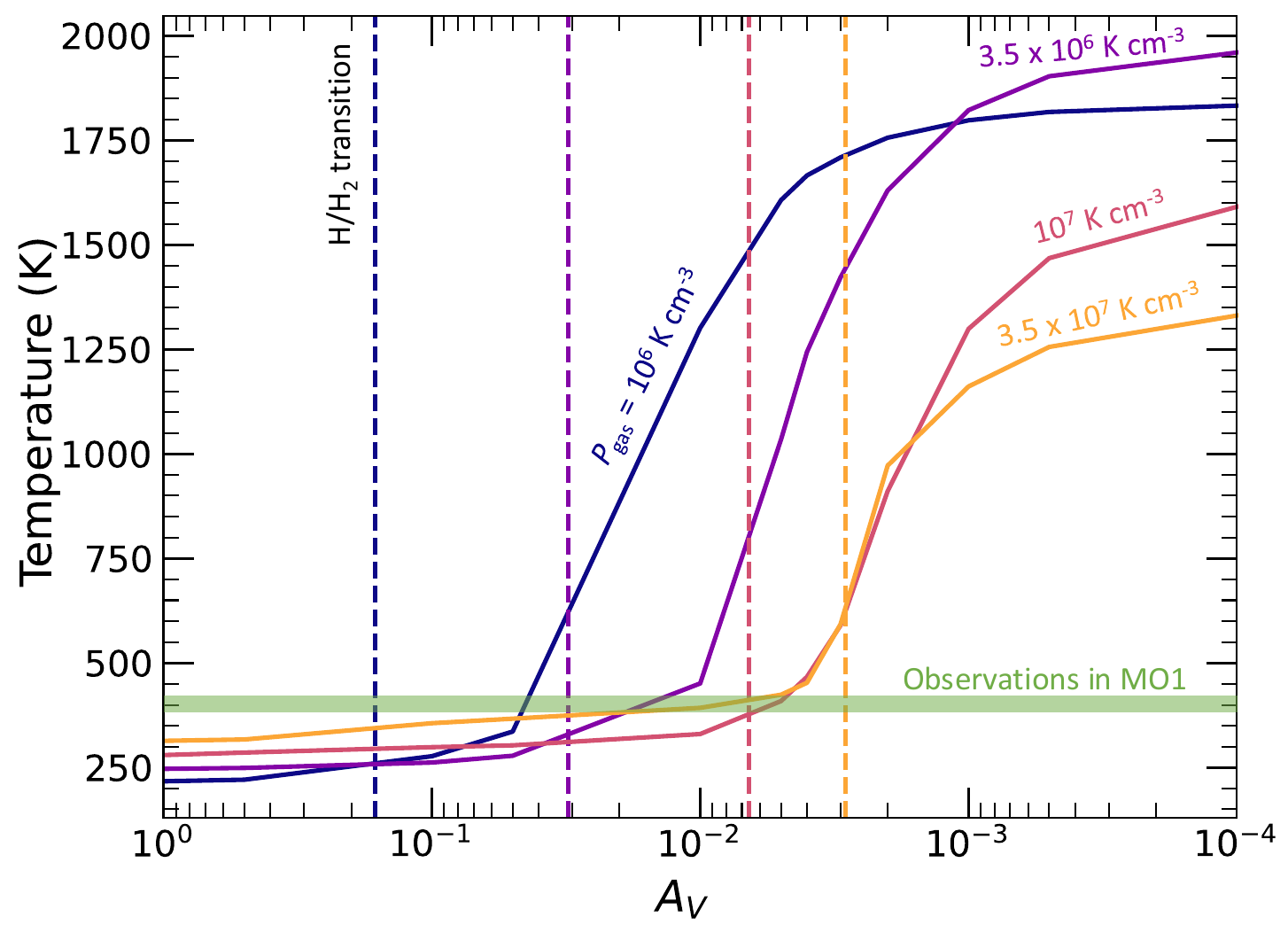}
    \includegraphics[width=\linewidth]{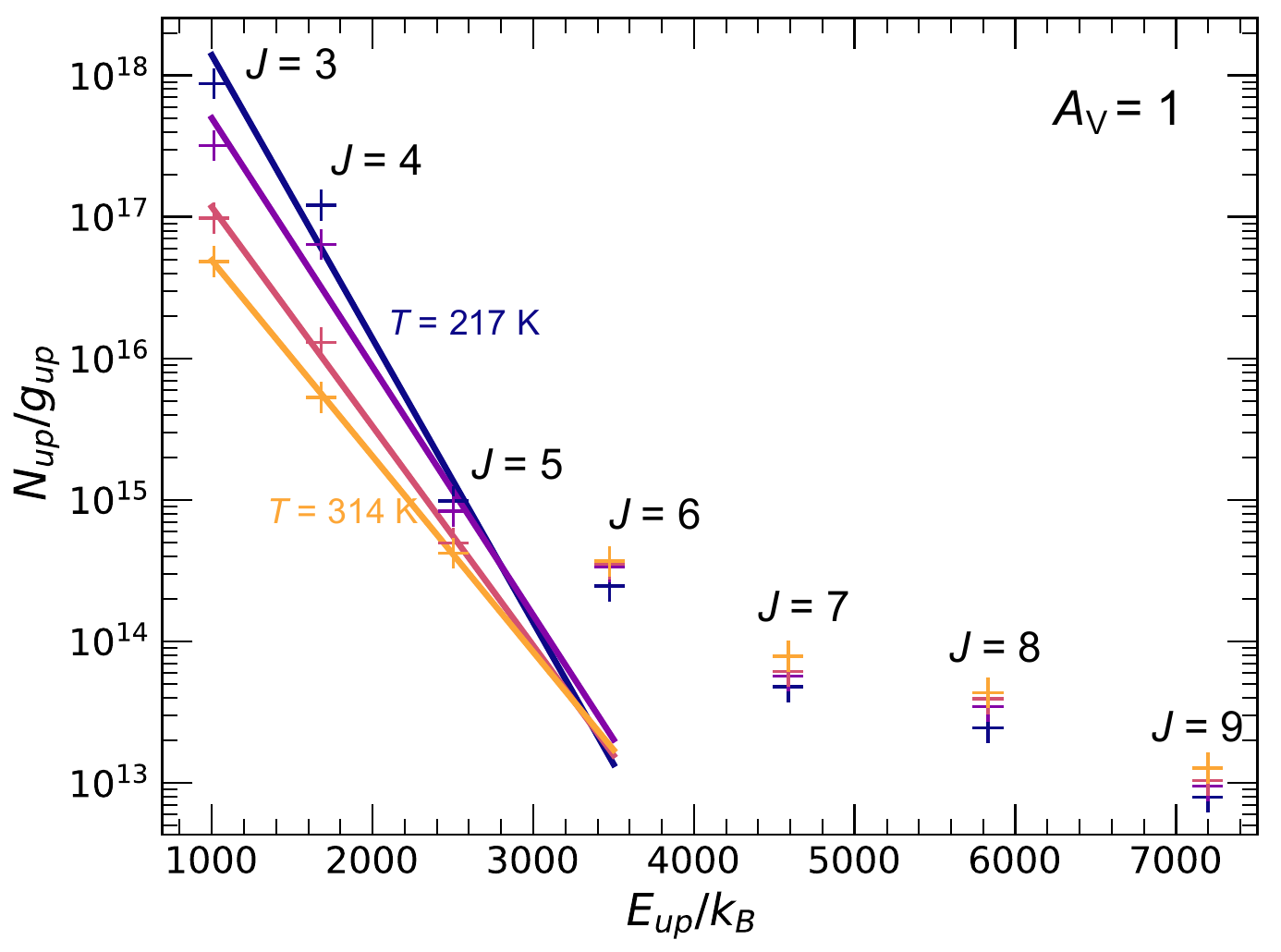}
    \includegraphics[width=\linewidth]{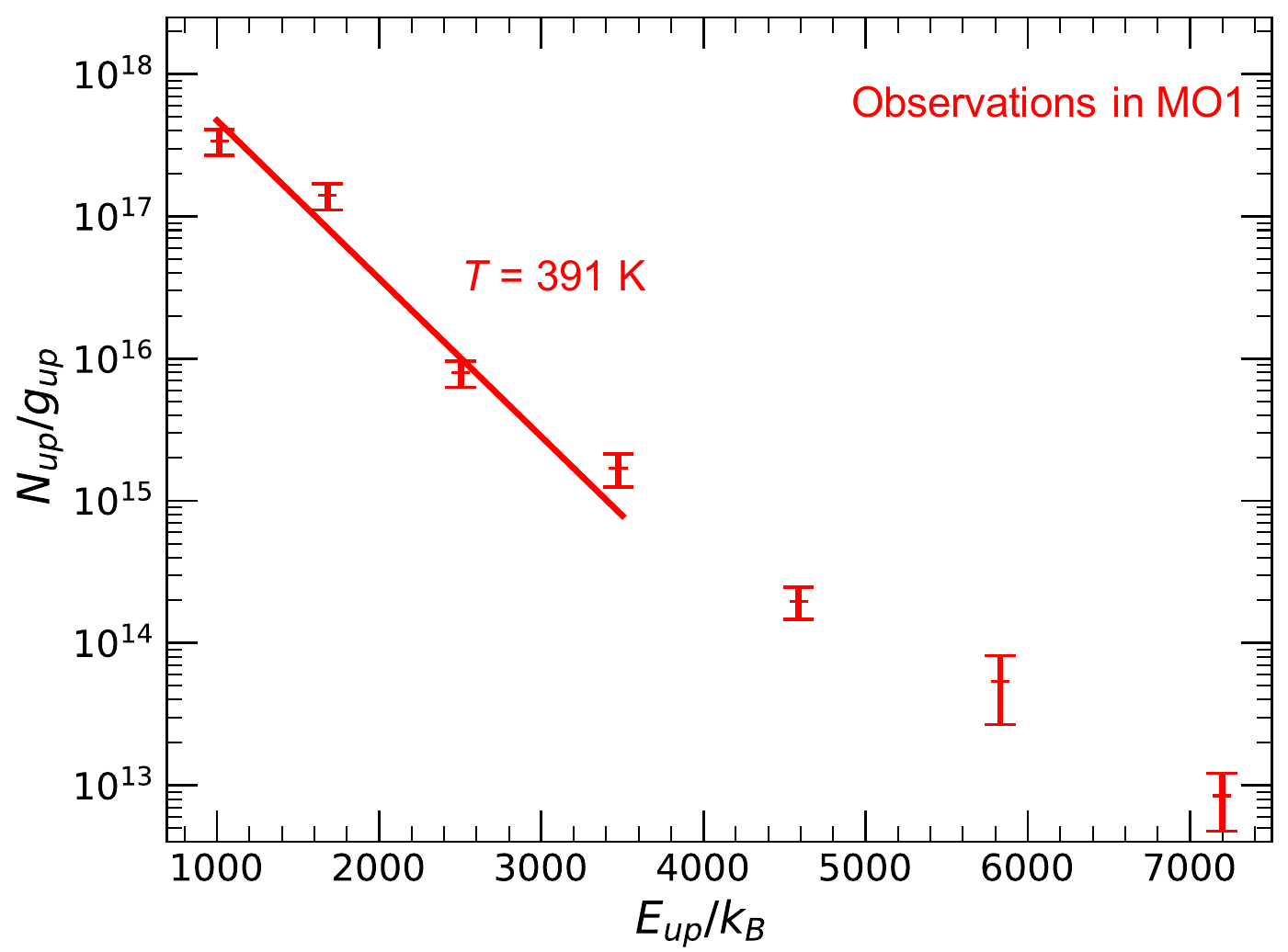}
    \caption{(Top) Excitation temperature derived from the first rotational lines of \Hmol\  as a function of $A_V$ for face-on isobaric stationary PDR models at different thermal pressure. (Middle) Excitation diagrams derived from the same models as the top panel at a position $A_V = 1$. (Bottom)  Excitation diagram observed in MO1, the region behind DF1.}
    \label{fig:models_temp_h2}
\end{figure}

Figure \ref{fig:models_temp_h2} displays the excitation temperature of \Hmol\ rotational lines (0--0 S(1) to S(3)) for isobaric PDR models when observed face-on at different thermal pressures. Observing face-on means that the observer is at a 0$\degree$ angle from the normal to the PDR. This amounts to integrating the intensity of the lines over the entire PDR ($A_V = 0-10$). Above $A_V = 0.5$, the \Hmol\ excitation temperature becomes constant close to the gas temperature expected at the H/\Hmol\ transition (see Fig. \ref{fig:nH_T}). Thus, we recover the fact that all the \Hmol\ emission originates from a thin layer at the edge of the PDR (so before $A_V = 0.5$). \Hmol\ lines are thus a good tracer of the gas temperature in the \Hmol\ emitting region. This figure shows a tendency where the higher the pressure is, the higher the gas temperature at the H/\Hmol\ transition is. In fact, the temperature derived from a model at $P_{\rm th} = 10^6$ K \cc\ is around $T \sim 200$ K when the observed temperature is always higher than 350 K. This result highlights the need for high pressure to reproduce the observed temperature with stationary models. 

However, even a high-pressure stationary model cannot reproduce the observed excitation diagram. The bottom panel of Fig. \ref{fig:models_temp_h2} displays the excitation diagram in the MO1 region (see Fig. \ref{fig:footpints_MRS}), which is after the 0--0 S(1) peak, around 5" from the edge and in which the line of sight should cross the entire PDR (and not only the illuminated edge). The observed temperature in MO1 can thus be compared to the temperature obtained from modeled \Hmol\ lines observed face-on. The temperature observed in this region, around $T_{\rm obs} = 391$ K, is always higher than the temperature derived in models, even at the highest thermal pressure $P_{\rm th} = 3.5 \times 10^7$ K \cc\ which is around $T_{\rm gas} = 314$ K. 
\footnotetext[5]{This parameter must not be confused with what is displayed in Fig. \ref{fig:Av_cut} which is the attenuation by the foreground matter; between \Hmol\ emitting region and the observer.} 
\footnotetext[6]{$d = \frac{N_\mathrm{H}}{E_{\mathrm{B-V}}}\, \frac{1}{R_\mathrm{V}}\, \int \frac{dA_\mathrm{V}}{n_\mathrm{H}\left( A_\mathrm{V} \right) }$, with $R_V = 3.1$} 
In addition, the overall shape of the excitation diagram is very different. In the observations, \Hmol\ column densities are aligned at least until the $J=7$ level, meaning that the Boltzmann law at the cold temperature can account for the population distribution for those levels.
In the model, the cold temperature can only account for the levels $J \leq 5$. This break in the excitation diagram is due to the difference in excitation mechanisms. Fig. 8.20 of \cite{maillard_model_2023} shows that for a model at $P_{\rm gas} = 4\times10^6$ K \cc\ and $G_0 = 100$, the excitation by collisions only dominates for levels $J\leq 5$. For more excited levels, the excitation is dominated by the IR cascade following UV pumping. The fact that a model with appropriate parameters for the Horsehead nebula cannot reproduce the observations has already been suggested by the previous \textit{Spitzer} observations of low/moderate excited PDRs \citep{habart_excitation_2011}. This preliminary study tends to unveil that a heating term, which would enhance the importance of collisions to excite higher energy levels, could be missing in the modeling. This missing heating term could be due to dynamical effects which are not considered in the Meudon PDR code or microphysics processes which are underestimated. We discuss these effects in the next sections. Further modeling is necessary to conclude on this matter and is out of the scope of this paper.

\subsection{Heating processes in stationary PDR models}

\begin{figure}
    \centering
    \includegraphics[width=1\linewidth]{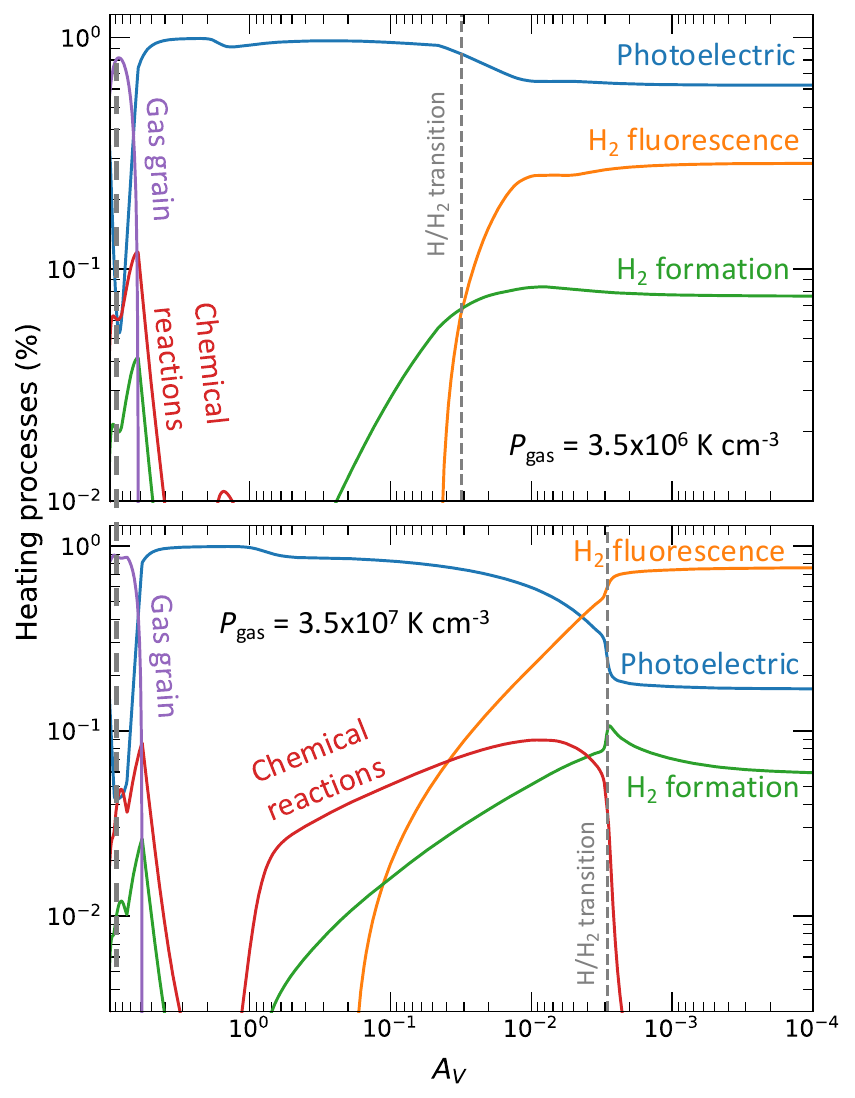}
    \caption{Main heating mechanisms in PDR models at (top) $P_{\rm gas} = 3.5 \times 10^6$ K cm$^{-3}$ (bottom) $P_{\rm gas} = 3.5 \times 10^7$ K cm$^{-3}$ and $G_0 = 100$.}
    \label{fig:heating_processes}
\end{figure}

Figure \ref{fig:heating_processes} displays the main heating processes in the PDR models at two different thermal pressures. This figure shows that in both models, the heating processes that dominate at the H/\Hmol\ transition are the photoelectric effect on the surfaces of dust grains and the radiative cascade of \Hmol. In higher pressure models, the radiative cascade of \Hmol\ dominates over photoelectric heating before the H/\Hmol\ transition. This might seem counterintuitive as the abundance of \Hmol\ is very low in this region. However, the dissociating efficiency of a photon is about $10\%$. Hence, the absorption of a UV photon leads, nine times out of ten, to a radiative cascade to an excited vibrational state. Then, de-excitation by collisions transfers kinetic energy into the medium, heating the gas. This mechanism is more efficient at high density (so at high pressure) because the collisions become competitive with the quadrupolar transitions in the NIR. 
After the H/\Hmol\ transition, the abundance of \Hmol\ increases rapidly, so \Hmol\ begins to self-shield and the radiative cascade is not efficient anymore. This mechanism is replaced by other processes. The dominant ones are the formation of \Hmol\ on dust grains, which releases kinetic energy into the medium and the heating by exothermic chemical reactions, which are increasingly important at high thermal pressure. Studying the importance of each process is essential to discuss what term could be underestimated and could lead to the discrepancy between observations and models.

The photoelectric effect is very efficient on small grains \citep[with radius $r_d < 10$ nm,][]{bakes_photoelectric_1994,weingartner_dust_2001,habart_photoelectric_2001}. Thus, the importance of this mechanism depends drastically on the abundance of nano-grains and their size distribution \citep[][Meshaka et al. 2024]{schirmer_influence_2021}. For instance, \cite{schirmer_influence_2021} show that nanograin-depleted regions could have lower gas temperatures compared to regions with ISM-like dust. 
The evolution of small dust grains at the edge of PDRs due to the intense UV field could have a great impact on the thermal balance. This could lead to models overestimating or underestimating the heating due to the photoelectric effect. Evaluating the dust size distribution is highly challenging. 
\citep{elyajouri_jwst_2025} have found, using JWST data, a similar nano-grain minimum size compared to the diffuse ISM but a less steep grain size distribution. Hence, the uncertainty on the extinction curve is a source of uncertainty on the thermal balance, as it influences the location of the H/\Hmol\ transition and the local gas temperature at this position.

In addition, the formation of \Hmol\ on the surfaces of dust grains is also badly constrained \citep[e.g.,][]{habart_empirical_2004,wakelam_h_2017} and could be a source of underestimation of the gas temperature at the H/\Hmol\ transition. In models,  equipartition is considered, so that only a third of the energy released from the reaction is transformed into kinetic energy. This heating process could be more important if the distribution of energy does not follow equipartition. Moreover, the formation rate of \Hmol\ is also badly constrained in PDRs since it is a function of the surface area of the small grains. The underestimation of the formation rate, which controls the location of the H/\Hmol\ transition zone, could move this transition closer to the PDR edge, where the gas temperature is higher. However, in the low/moderately excited PDR regime, with low $G_0$/$n_{\rm H}$ ratios, \Hmol\ self-shields efficiently enough so that the H/\Hmol\ transition zone is already closer to the edge.

\subsection{Photoevaporation-induced dynamical effects on \Hmol\ excitation}
\label{subsec:dynamics}

In the following, we discuss the possibility that the highly-pressurized and highly-excited \Hmol\ detected at the edge of the PDR could originate from dynamical effects at the interface between the neutral and ionized media. At the edge of the PDR, imaging data reveal a regular pattern of "fingers-like" structures, which look like small-scale cometary globules seen in photoevaporating molecular clouds \citep[see Fig.~\ref{fig:footpints_MRS} and ][]{bertoldi_photoevaporation_1990, lefloch_cometary_1994}. Our estimate of the thermal pressure\footnote{The total pressure (including magnetic and turbulent) is higher than the thermal pressure: $P_{\rm tot} = P_{\rm th} + P_{\rm B} + P_{\rm turb}$. The magnetic pressure can be estimated as $P_{\rm B} = \frac{B^2}{8 \pi} \sim 9\times 10^5$ K cm$^{-3}$ with $B = 56$ $\mu$G \citep{hwang_magnetic_2023}. The turbulent pressure can be estimated as $P_{\rm turb} = \mu n m_{\rm H} \sigma_v^2 \sim 10^6$ K cm$^{-3}$ where $m_{\rm H}$ is the mass of the hydrogen atom, $\mu \sim 2$ is the mean molecular weight, the gas density is $n = 10^4$ cm$^{-3}$ and the velocity dispersion $\sigma_v \sim 1$ km s$^{-1}$. The turbulent pressure is not negligible in the PDR, as it can reach the same order of magnitude as the thermal pressure. However, given that the \Hmol\ lines are not spectrally resolved, it is impossible to address the nature of the broadening. We therefore only consider the comparison of the estimates of the thermal pressures in the molecular and ionized regions.}  from \Hmol\ lines shows that the PDR front is at higher pressure than the \Hii\ region, which suggests that dynamical effects are very important where the intensity of the \Hmol\ lines rise steeply, within around 1" from the ionizing front. This 1"-scale matches the "fingers-like" structures seen in the imaging data, perpendicular to the interface, and we propose that this corresponds to the scale at which thermal instability develops because of mixing induced by the photo-evaporation flow. 

The photo-evaporation gas flow at the edge of the PDR triggers advection and mixing between the cold neutral medium (CNM) and the more diffuse (neutral and ionized) medium, which creates turbulent and thermally unstable gas \citep{bertoldi_nonequilibrium_1996, nakatani_photoevaporation_2019}. 3D numerical simulations of turbulent ISM show that the mixing across cold and more diffuse gas phases leads to warm and out-of-equilibrium \Hmol\, initially formed in the CNM and injected in more diffuse environments \citep{valdivia_h2_2016, bellomi_3d_2020, godard_3d_2023}, like the warm neutral medium (WNM), which could be responsible for the highly-excited \Hmol\ we detect at the front. As the temperature in the ionized region, $T_e \sim 8000$ K, is significantly higher than the temperature in the PDR, this mixing is likely to result in high temperatures at the PDR interface. Hence, this could explain why the observed temperature is higher than what is predicted by 1D stationary PDR models even at high thermal pressure. The constraint of the impact of photoevaporation on the thermal balance and the formation of these small-scale structures requires proper numerical modeling with a dynamical and thermochemical code, which is beyond the scope of this paper.

Another argument for the importance of the dynamical effects is that the ionization front and the dissociation front are not clearly spatially resolved and could possibly be merged (H/\Hmol\ - IF < 100 au), even in the high spatial resolution data of the JWST. \cite{maillard_dynamical_2021} shows that dynamical effects can significantly reduce the size of atomic layers and even lead to merged fronts. These effects are expected to be particularly important for low excitation PDRs such as the Horsehead. Indeed, for PDRs with low $G_0$/$n_{\rm H}$, the \Hmol\ self-shielding is the dominant source of FUV absorption compared to dust extinction \citep{sternberg_h_2014}. This means that the advection of the ionization front has a stronger effect on the merging of the fronts because \Hmol\ can still be shielded from the UV field. Hence, lower values of the advection velocity are needed to merge the fronts.

\section{Conclusions}

In this work, we have studied the spatial morphology and excitation of \Hmol\ in the Horsehead nebula. We used \Hmol\ lines to constrain the physical parameters, such as the gas temperature and thermal pressure, of the PDR.
The main conclusions of this study can be summarized as follows:

\begin{enumerate}
    \item We reveal the first spatial separation between \Hmol\ lines peaks around $d \sim 0.5$". FUV-pumped lines ($v=0$ $J_u>6$ and $v>0$) peak closer to the edge than collisionally excited lines (0--0 S(1)-S(4)). The emission of AIBs is more correlated to the low-excitation \Hmol\ lines than the FUV-pumped lines. 
    
    \item We derived the attenuation profile across the PDR using rovibrational \Hmol\ lines in the near-infrared. The attenuation increases from $A_V = 0.5$ at the edge to $A_V = 8$ at 12" from the edge, considering $R_V = 3.1$. This is in agreement with a geometry where the Horsehead is illuminated on its backside and DF2 is located further away from the observer than DF1.
    
    \item The measure of the \Hmol\ column density reveals that the column density toward DF2 ($N(\Hmol) = 1.9 \times 10^{20}$ cm$^{-2}$) is 5 times higher than in DF1 ($N(\Hmol) = 3.8 \times 10^{19}$ cm$^{-2}$).
    \item The observed temperature derived from \Hmol\ lines does not vary much throughout the PDR. It decreases from 550 K at the edge to 380 K inside the PDR. The Horsehead not being observed exactly edge-on could explain this small observed variation. If there is illuminated material all over the field of view, then \Hmol\ lines trace the temperature in the illuminated layer and are not a good tracer of the gas temperature inside the PDR. 
    
    \item The OPR is never at equilibrium over the field of view. Its value varies from OPR $\sim$ 2-2.5 at the edge of each dissociation to OPR $\sim$ 1.3-1.5 deeper inside the PDR. Its variations follow roughly the variation of the observed temperature: the OPR reaches its highest value where the temperature is higher. 
    
    \item We use the \Hmol\ emission at the very edge of the PDR ($d < 1.5$"), where all the gas is illuminated, to derive an estimate of the thermal pressure. We derive a lower limit value around $P_{\rm gas} \geq 6 \times 10^6$ K \cc. This value is higher than the thermal pressure predicted in the \Hii region. 

    \item Template stationary 1D PDR models cannot account for the intrinsic 2D structure and the very high temperatures observed in the Horsehead nebula, which has already been suggested by the previous \textit{Spitzer} \citep{habart_excitation_2011}. This preliminary study suggests that a heating term is missing in the modeling.

    \item We propose that extra heating could originate from the mixing between the molecular and the more diffuse atomic/ionized gas at the edge of the PDR, induced by the photoevaporation of the cloud. Detailed modeling is needed to link the highly excited and over-pressurized \Hmol\ detected in the illuminated filaments at the interface.   

\end{enumerate}

In conclusion, our study has revealed the complexity of observing cold \Hmol\, as its emission is always dominated by the illuminated matter present in the field of view. Despite this result, we have been able to derive a very high pressure at the edge of the PDR, which explains the photoevaporating flow observed in the imaging data. This observed high pressure implies that the dynamical effects are probably not negligible in this region. Dynamical effects are probably the reason for the very small size of the atomic layer, with maybe merged ionized and dissociation fronts, and explain why stationary PDR models cannot reproduce \Hmol\ excitation. Dynamical thermochemical models, such as Hydra \citep{bron_photoevaporating_2018}, are necessary to constrain the impact of dynamics on the thermal balance and \Hmol\ excitation in the Horsehead. To complement the study of dynamics, a detailed analysis of \Hmol\ excitation and ortho-para conversion, with state-of-the-art PDR models, is necessary. This will allow the constraint of specific mechanisms (ortho-para conversion in the gas and at the dust surface, different self-shielding in ortho and para levels, possible excitation by cosmic rays in the inner PDR...). Deeper observations are needed to better constrain the emission in deeper layers of the PDR, such as the molecular region.

\begin{appendix}
    
\onecolumn
\section{Comparison between imaging (NIRCam, MIRIm) and spectro-imaging (NIRSpec, MIRI-MRS)}
\begin{figure*}[!h]
  \includegraphics[width=0.49\textwidth]{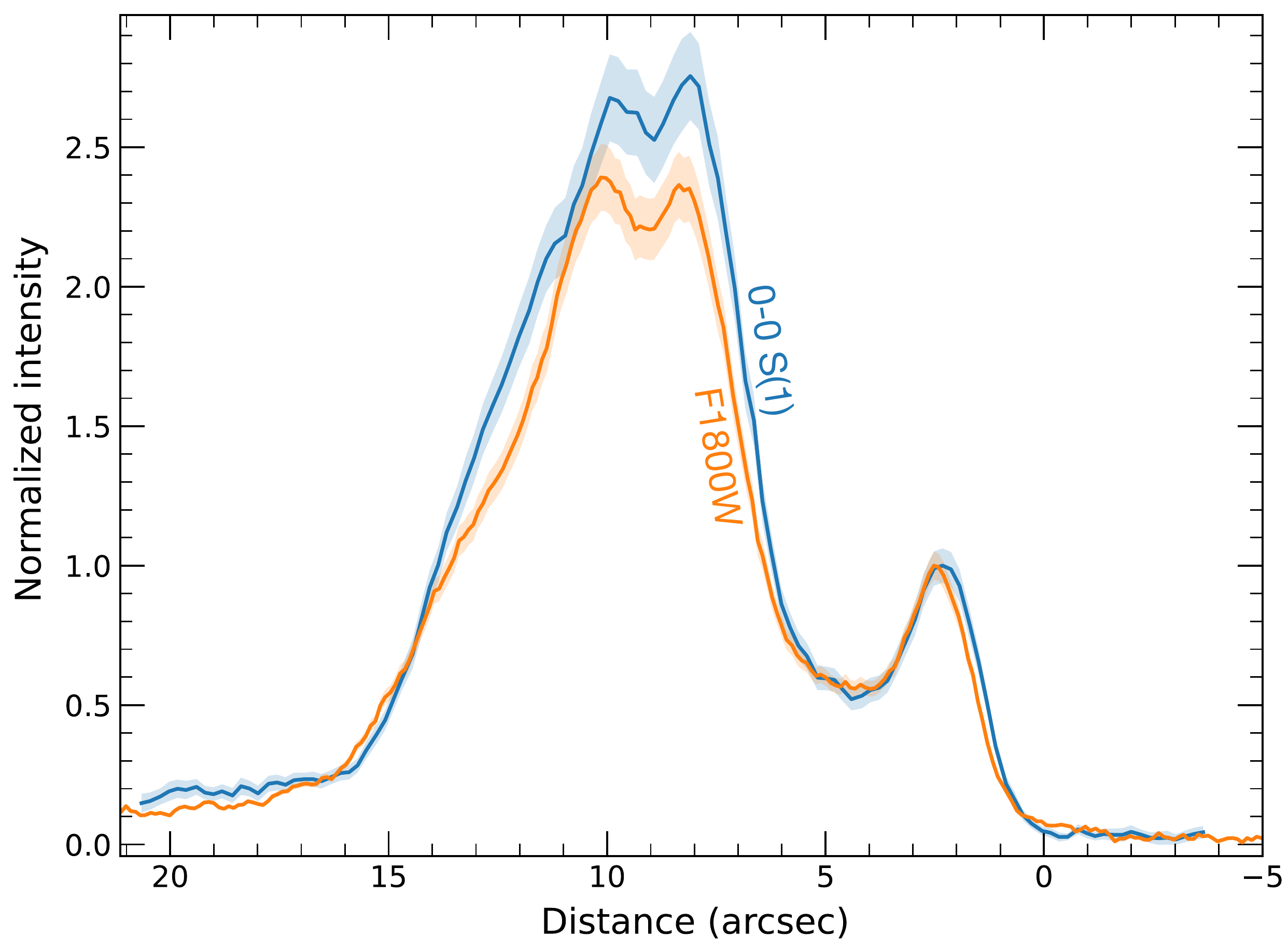}
  \includegraphics[width=0.49\textwidth]{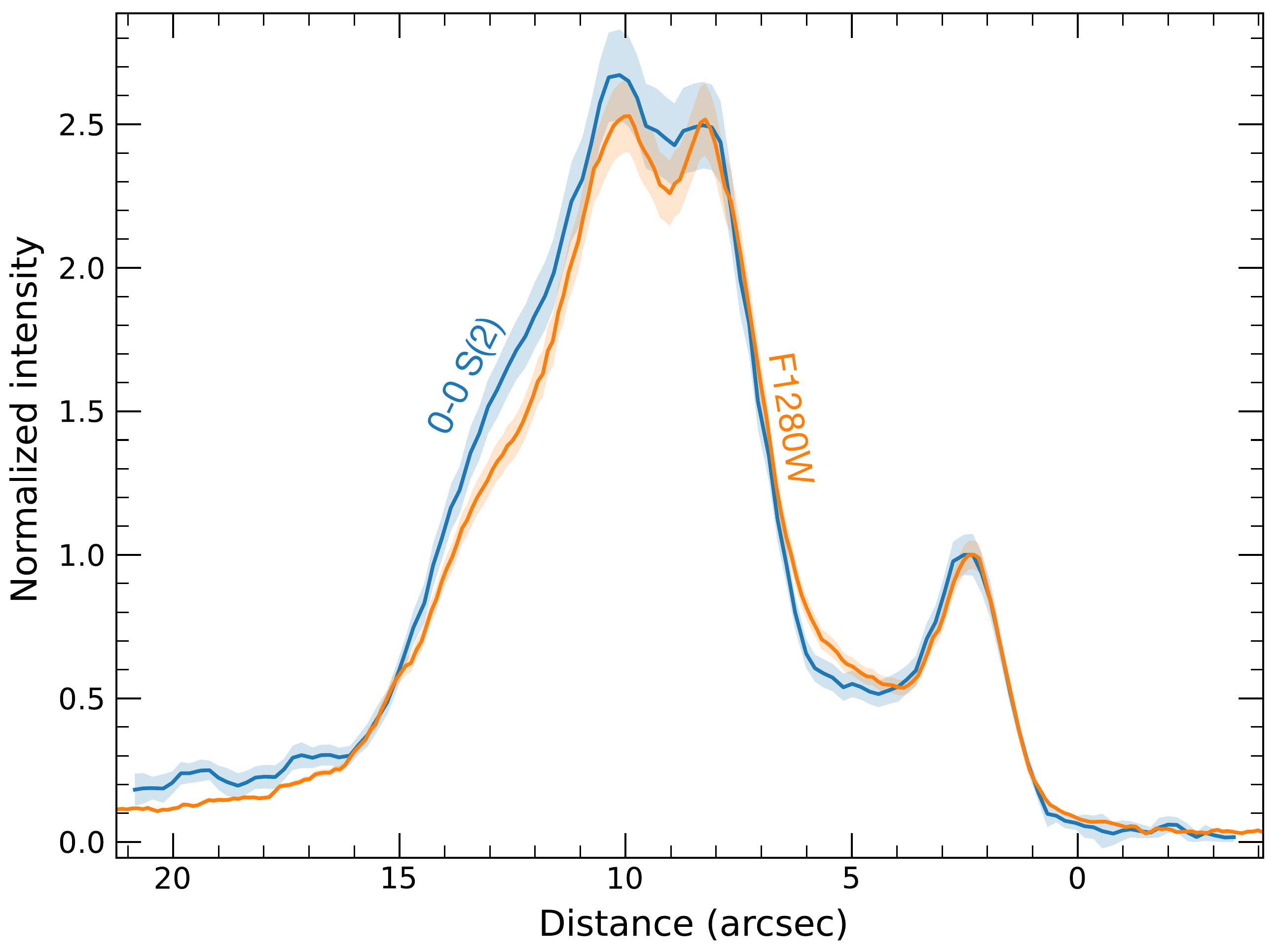}
  \includegraphics[width=0.49\textwidth]{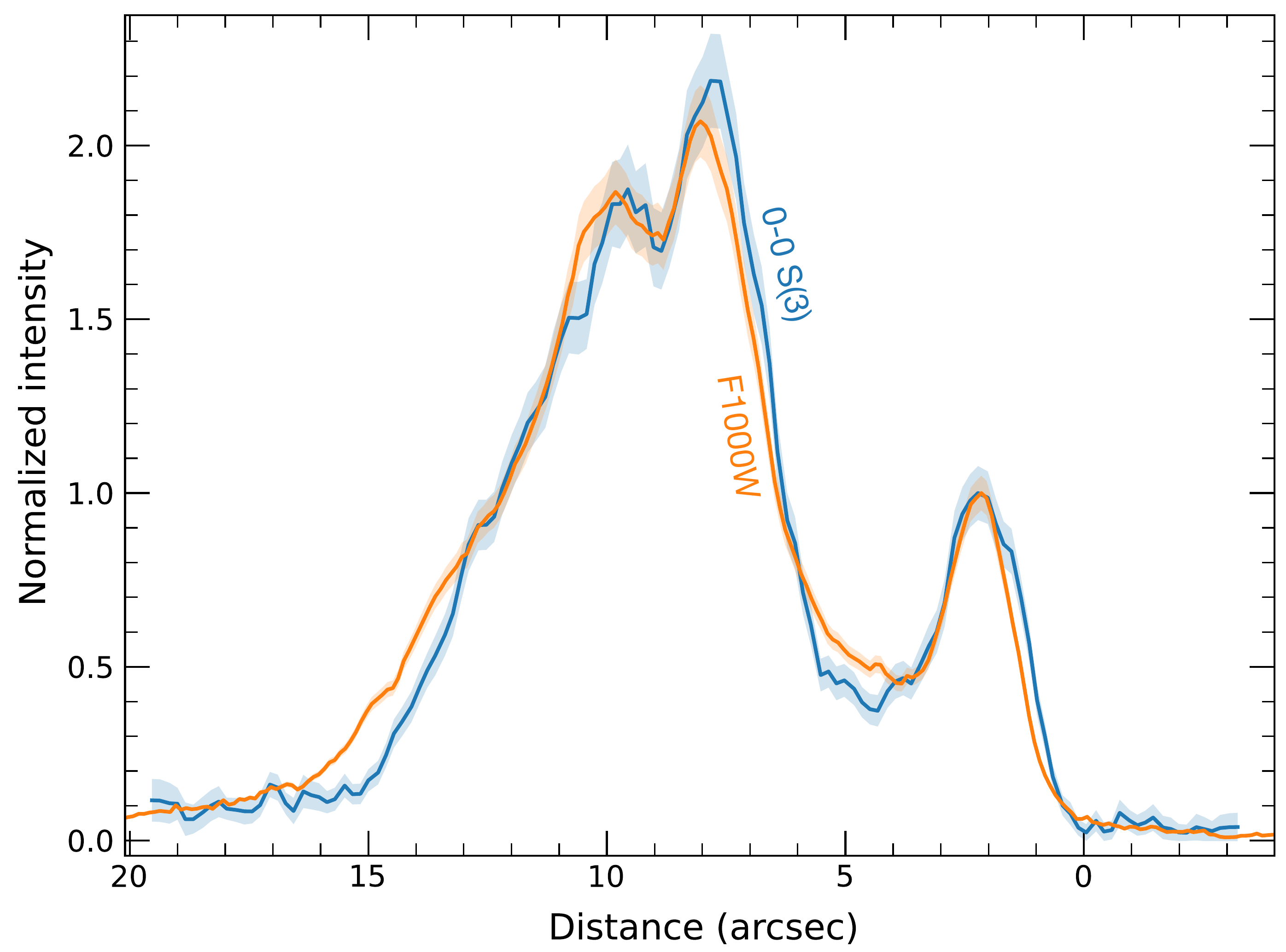}
  \includegraphics[width=0.49\textwidth]{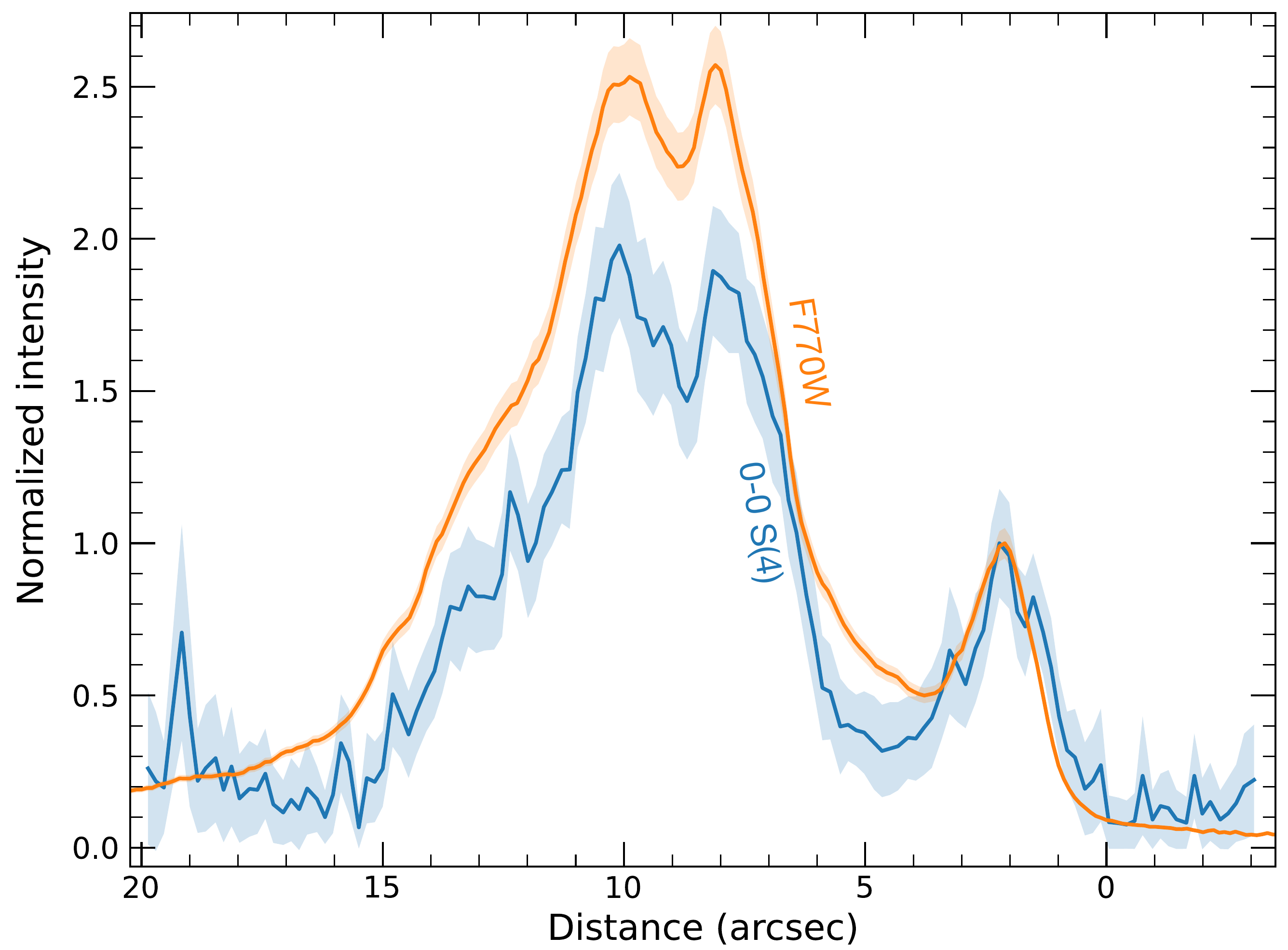}
 \caption{Comparison of \Hmol\ lines and dust emission profiles from imaging data across the front \citep[cut \#3 from][]{abergel_jwst_2024} averaged on 0.5" perpendicular to the line cut. Profiles are flux-normalized between 0.5 and 3" around the first peak.}
\label{fig:H2_H2_plus_dust_profiles}
\end{figure*}

\begin{figure*}[!h]
  \includegraphics[width=0.49\textwidth]{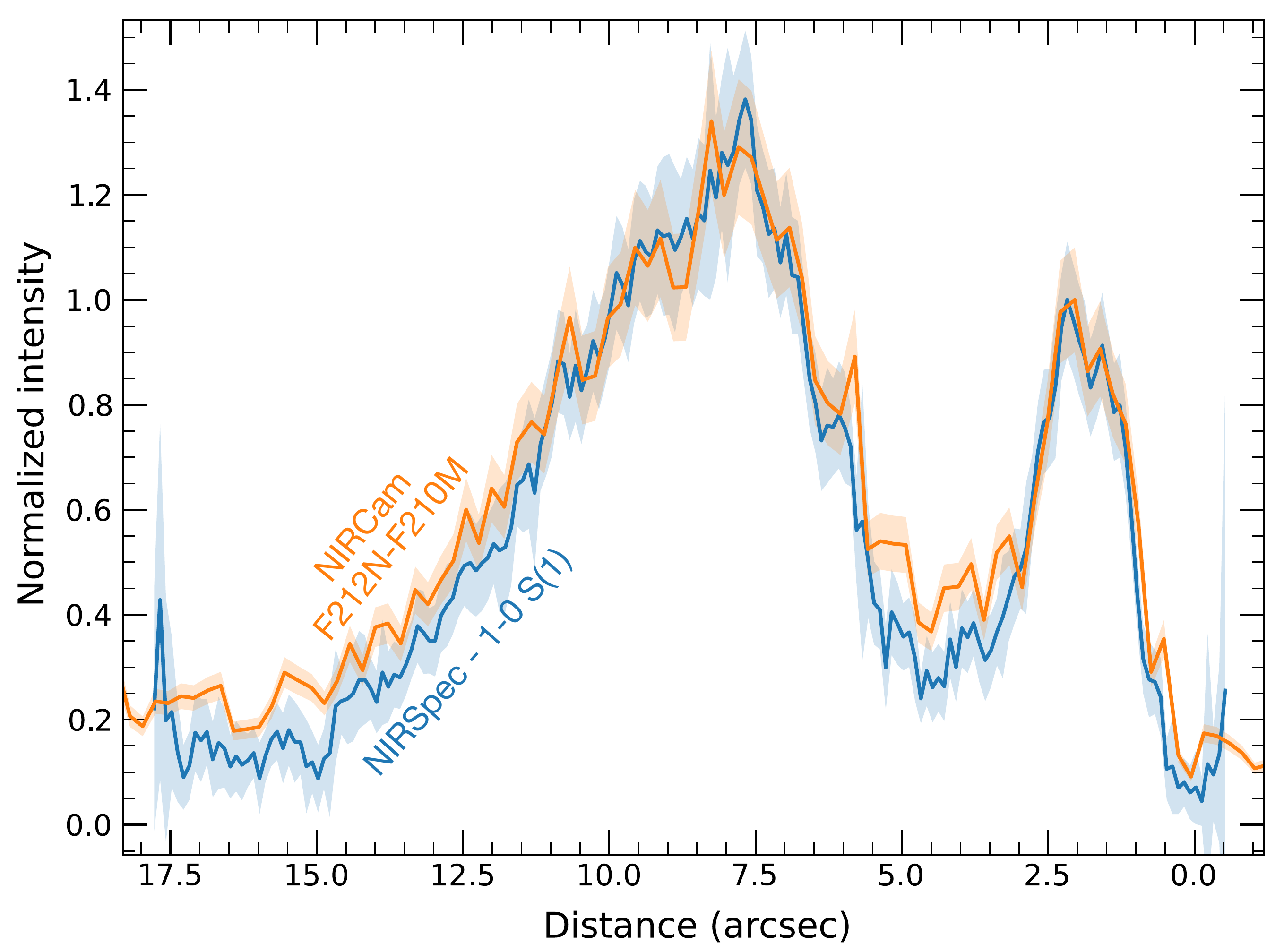}
  \includegraphics[width=0.49\textwidth]{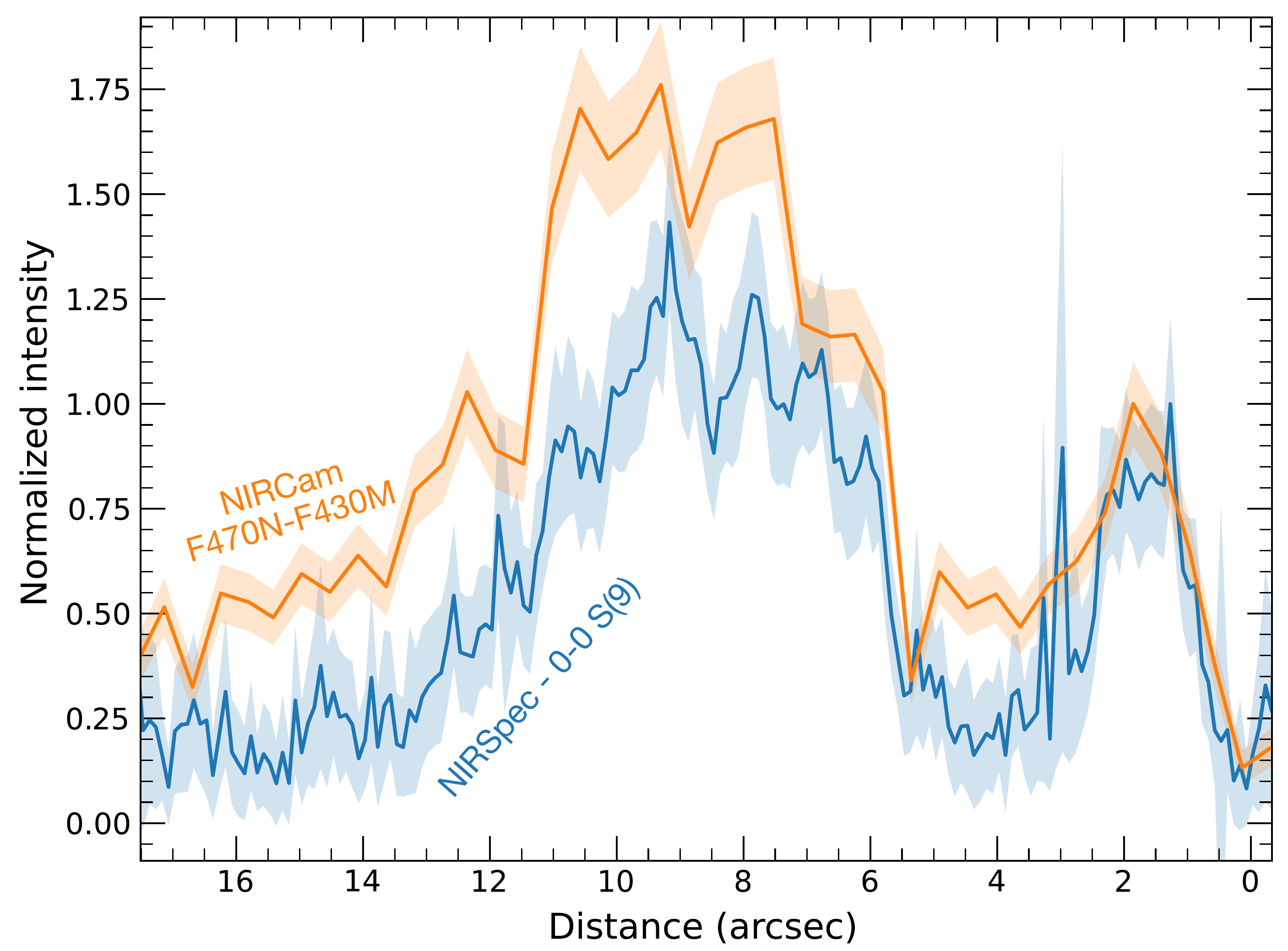}
 \caption{Comparison of \Hmol\ lines and emission profiles from imaging data across the front \citep[cut \#3 from][]{abergel_jwst_2024} averaged on 0.5" perpendicular to the line cut. Profiles are flux-normalized between 0.5" and 3" (average around the first peak). 
  }
\label{fig:H2_H2_plus_dust_profiles}
\end{figure*}

\FloatBarrier

\twocolumn
\section{Full excitation diagram}
\label{app:excitation_diagram}
\begin{figure}[!h]
    \centering
    \includegraphics[width=\linewidth]{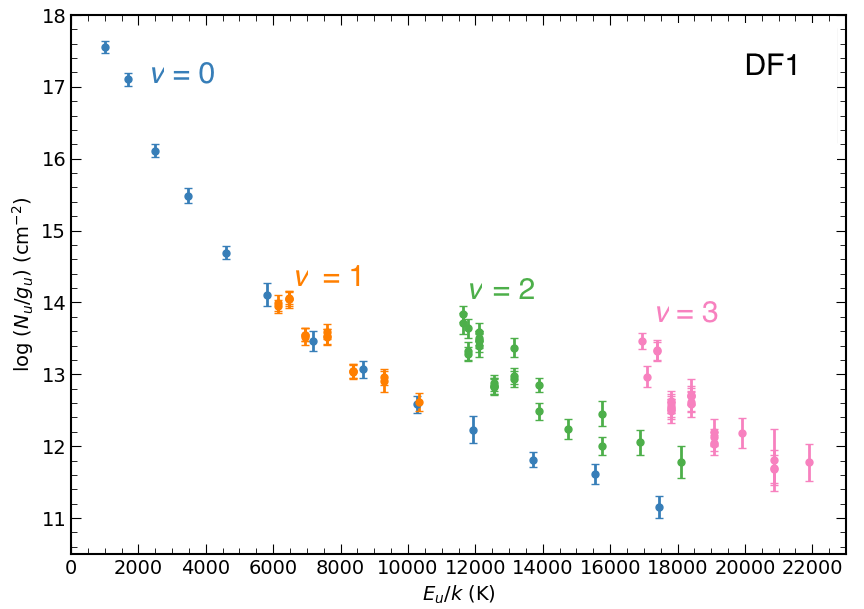}
    \includegraphics[width=\linewidth]{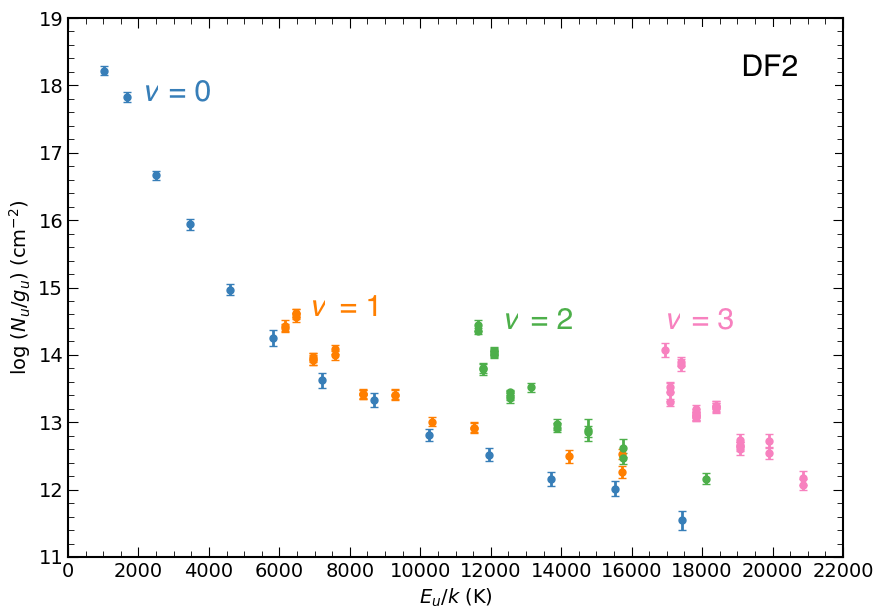}
    \caption{Total excitation diagram (top) in the DF1 (bottom) in the DF2, apertures defined in \cite{misselt_jwst_2025}.}
    \label{fig:diag_rot_tot}
\end{figure}

\section{Comparison of ortho and para line profile}
\begin{figure}[!h]
    \centering
    \includegraphics[width=\linewidth]{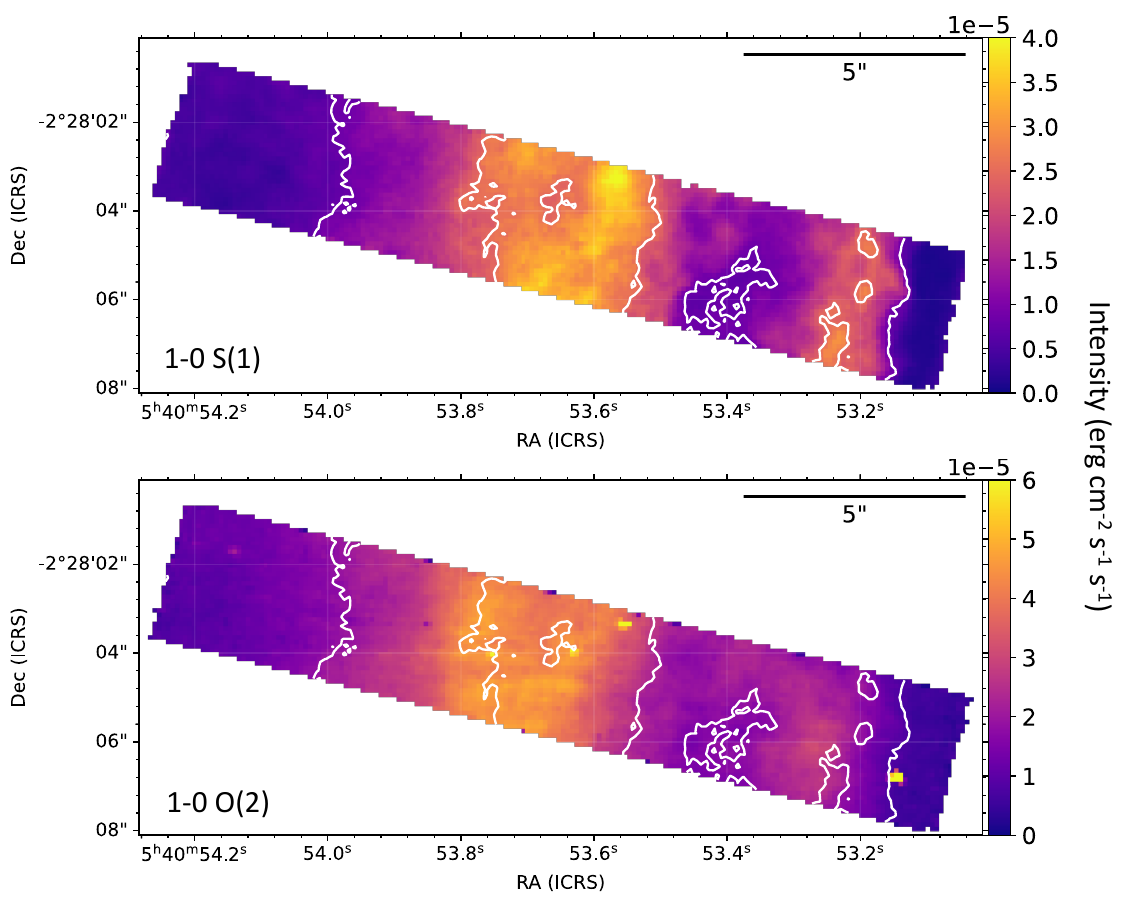}
    \caption{Comparison of the spatial distribution of an ortho rovibrational line (1--0 S(1)) and a para rovibrational line (1--0 O(2)). White contours are from the 1--0 S(1) line emission. A spatial shift is observed between these lines in the data.}
    \label{fig:map_10S1_10O2}
\end{figure}

\begin{figure}
    \centering
    \includegraphics[width=\linewidth]{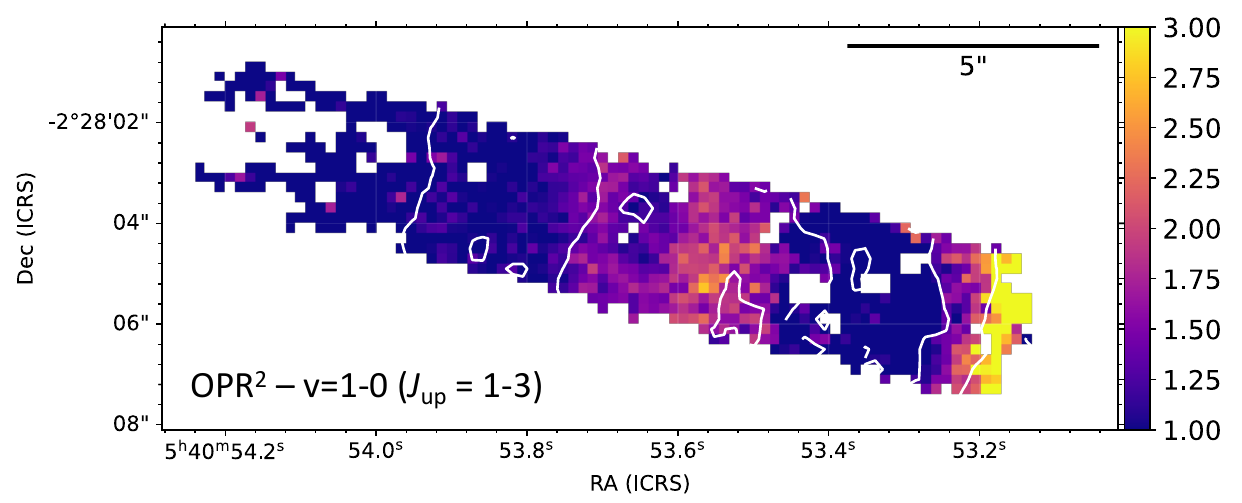}
    \caption{Ortho-to-para ratio squared map corrected for extinction calculated for $v=1-0$ $J_{\rm up} = 1-3$. White contours are the rotational OPR (levels: 1, 1.5, 2, see Fig. \ref{fig:map_temp_coldens}).}
    \label{fig:opr_rovib}
\end{figure}
\onecolumn
\begin{figure}[!h]
    \centering
    \includegraphics[width=0.9\linewidth]{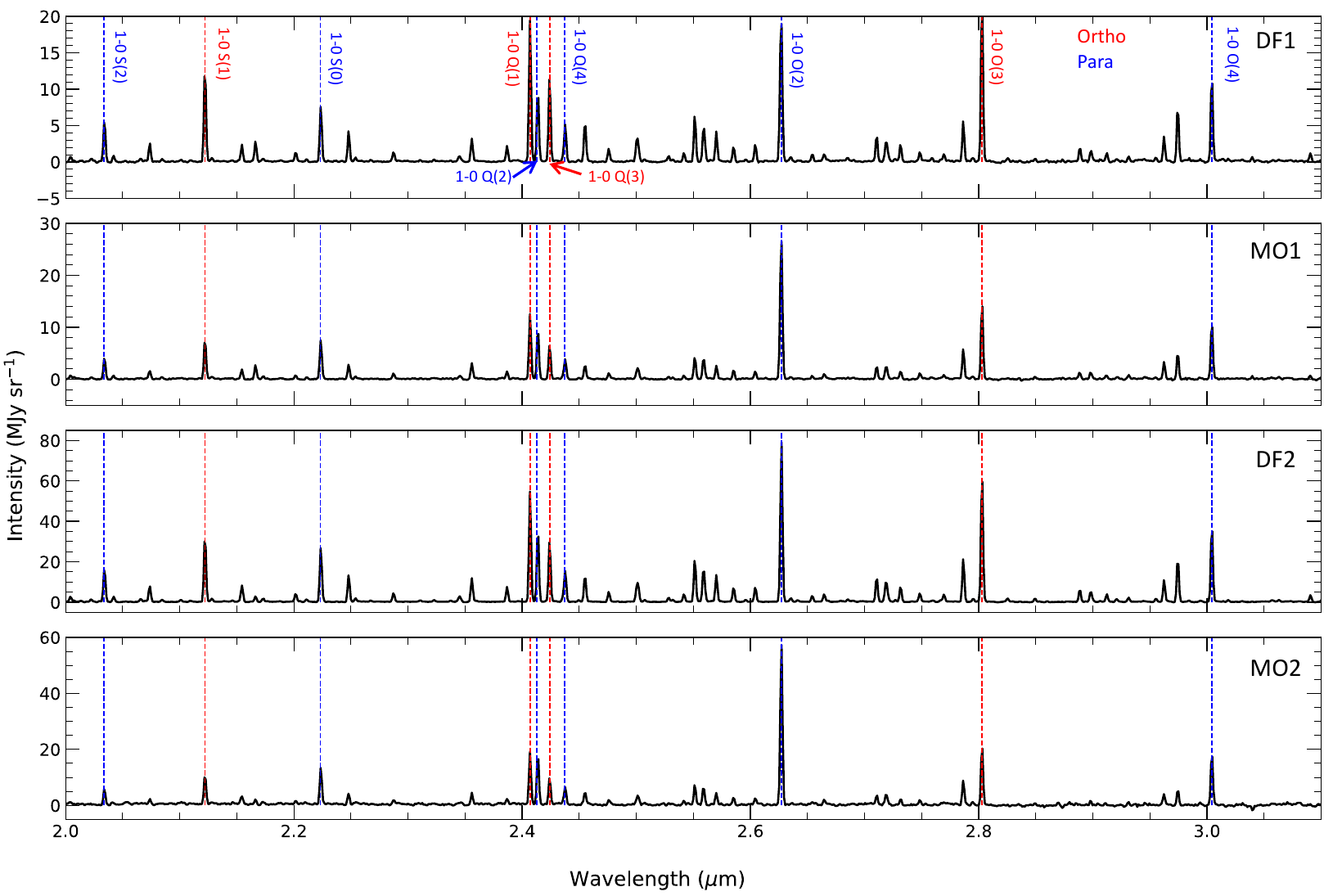}
    \caption{NIRSpec spectra corrected from extinction between 2 and 3.1 \mum\ in each of the template region defined in \cite{misselt_jwst_2025} (DF: dissociation front, MO: molecular region). The OPR is lower in the "molecular" regions than in the dissociation fronts.}
    \label{fig:spec_region}
\end{figure}
\begin{figure}[!h]
    \centering
    \includegraphics[width=0.9\linewidth]{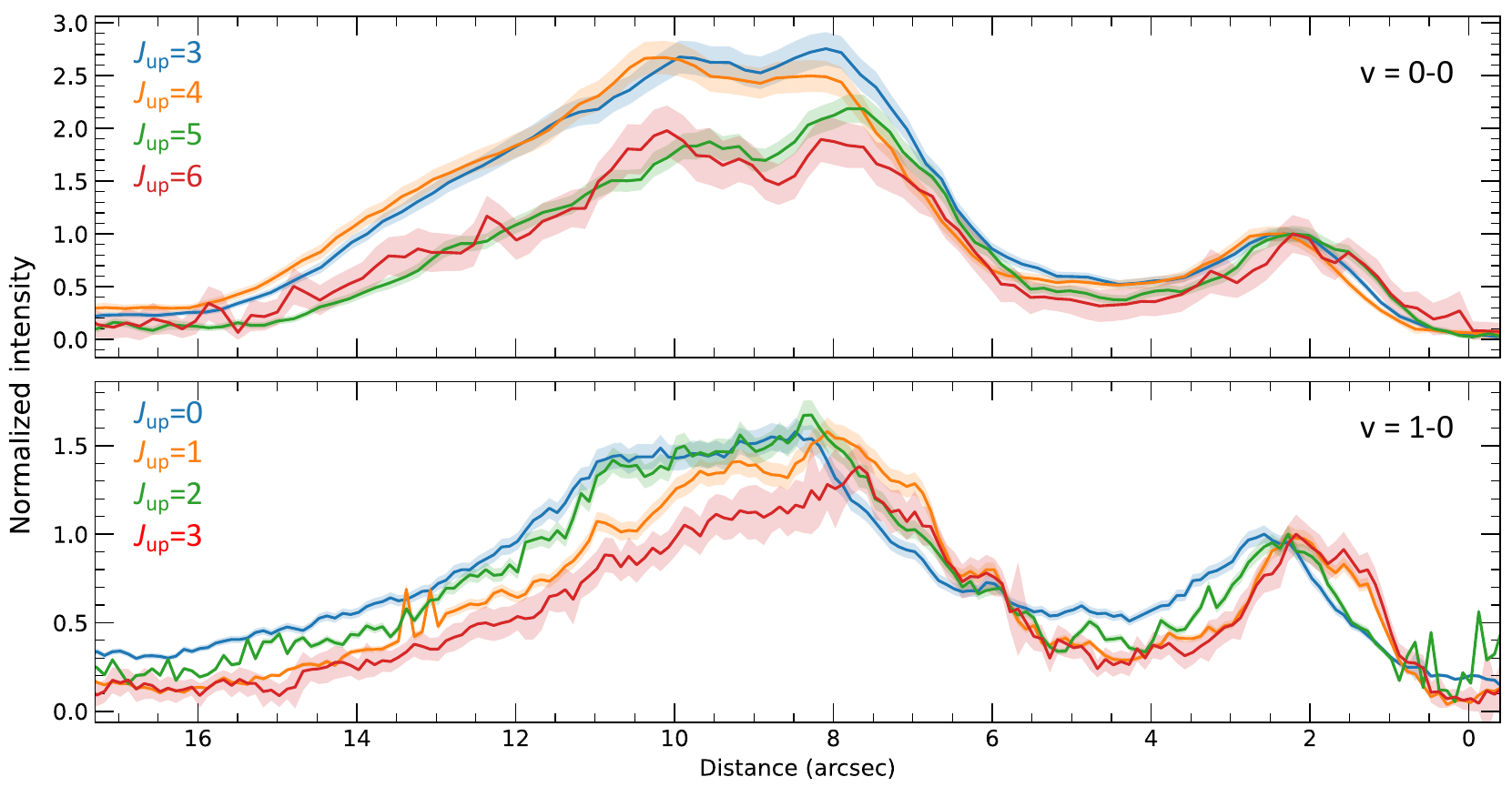}
    \caption{Normalized intensity (around 0.5 and 3" around the first peak) profiles across the front \citep[cut \#3][]{abergel_jwst_2024} averaged on 0.5" perpendicular to the line cut. The illuminating star is on the right. (Top) Observed first rotational levels. (Bottom) Observed first rovibrational levels. Para rovibrational levels peaks behind ortho rovibrational levels and the pure rotational para line 0--0 S(2) peaks behind the ortho line 0--0 S(1).}
    \label{fig:cut_ortho_para}
\end{figure}

\end{appendix}


\begin{thebibliography}{62}
	\expandafter\ifx\csname natexlab\endcsname\relax\def\natexlab#1{#1}\fi
	
	\bibitem[{Abergel {et~al.}(2024)Abergel, Misselt, Gordon, Noriega-Crespo,
		Guillard, Van De~Putte, Witt, Ysard, Baes, Beuther, Bouchet, Brandl,
		Elyajouri, Kannavou, Kendrew, Klassen, \& Trahin}]{abergel_jwst_2024}
	Abergel, A., Misselt, K., Gordon, K., {et~al.} 2024, A\&A
	
	\bibitem[{Anthony-Twarog(1982)}]{anthony-twarog_h-beta_1982}
	Anthony-Twarog, B.~J. 1982, AJ, 87, 1213
	
	\bibitem[{Argyriou {et~al.}(2023)Argyriou, Glasse, Law, Labiano, Álvarez
		Márquez, Patapis, Kavanagh, Gasman, Mueller, Larson, Vandenbussche, Glauser,
		Royer, Dicken, Harkett, Sargent, Engesser, Jones, Kendrew, Noriega-Crespo,
		Brandl, Rieke, Wright, Lee, \& Wells}]{argyriou_jwst_2023}
	Argyriou, I., Glasse, A., Law, D.~R., {et~al.} 2023, A\&A, 675, A111,
	publisher: EDP Sciences
	
	\bibitem[{Bakes \& Tielens(1994)}]{bakes_photoelectric_1994}
	Bakes, E. L.~O. \& Tielens, A. G. G.~M. 1994, ApJ, 427, 822
	
	\bibitem[{Bally {et~al.}(2018)Bally, Chambers, Guzman, Keto, Mookerjea,
		Sandell, Stanke, \& Zinnecker}]{bally_kinematics_2018}
	Bally, J., Chambers, E., Guzman, V., {et~al.} 2018, AJ, 155, 80
	
	\bibitem[{Bellomi {et~al.}(2020)Bellomi, Godard, Hennebelle, Valdivia,
		Pineau~des Forêts, Lesaffre, \& Pérault}]{bellomi_3d_2020}
	Bellomi, E., Godard, B., Hennebelle, P., {et~al.} 2020, A\&A, 643, A36, aDS
	Bibcode: 2020A\&A...643A..36B
	
	\bibitem[{Bertoldi \& Draine(1996)}]{bertoldi_nonequilibrium_1996}
	Bertoldi, F. \& Draine, B.~T. 1996, ApJ, 458, 222
	
	\bibitem[{Bertoldi \& McKee(1990)}]{bertoldi_photoevaporation_1990}
	Bertoldi, F. \& McKee, C.~F. 1990, ApJ, 354, 529, publisher: IOP ADS Bibcode:
	1990ApJ...354..529B
	
	\bibitem[{Bron {et~al.}(2018)Bron, Agúndez, Goicoechea, \&
		Cernicharo}]{bron_photoevaporating_2018}
	Bron, E., Agúndez, M., Goicoechea, J.~R., \& Cernicharo, J. 2018, ArXiv
	e-prints
	
	\bibitem[{Bron {et~al.}(2014)Bron, Le~Bourlot, \& Le~Petit}]{bron_surface_2014}
	Bron, E., Le~Bourlot, J., \& Le~Petit, F. 2014, A\&A, 569, A100, \_eprint:
	1407.4473
	
	\bibitem[{Bron {et~al.}(2016)Bron, Le~Petit, \&
		Le~Bourlot}]{bron_efficient_2016}
	Bron, E., Le~Petit, F., \& Le~Bourlot, J. 2016, A\&A, 588, A27
	
	\bibitem[{Bushouse {et~al.}(2023)Bushouse, Eisenhamer, Dencheva, Davies,
		Greenfield, Morrison, Hodge, Simon, Grumm, Droettboom, Slavich, Sosey, Pauly,
		Miller, Jedrzejewski, Hack, Davis, Crawford, Law, Gordon, Regan, Cara,
		MacDonald, Bradley, Shanahan, Jamieson, Teodoro, \&
		Williams}]{bushouse_jwst_2023}
	Bushouse, H., Eisenhamer, J., Dencheva, N., {et~al.} 2023, {JWST} {Calibration}
	{Pipeline}
	
	\bibitem[{Compiègne {et~al.}(2007)Compiègne, Abergel, Verstraete, Reach,
		Habart, Smith, Boulanger, \& Joblin}]{compiegne_aromatic_2007}
	Compiègne, M., Abergel, A., Verstraete, L., {et~al.} 2007, A\&A, 471, 205
	
	\bibitem[{de~Boer(1983)}]{de_boer_diffuse_1983}
	de~Boer, K.~S. 1983, A\&A, 125, 258, aDS Bibcode: 1983A\&A...125..258D
	
	\bibitem[{Decleir {et~al.}(2022)Decleir, Gordon, Andrews, Clayton, Cushing,
		Misselt, Pendleton, Rayner, Vacca, \& Whittet}]{decleir_spex_2022}
	Decleir, M., Gordon, K.~D., Andrews, J.~E., {et~al.} 2022, ApJ, 930, 15
	
	\bibitem[{Elyajouri {et~al.}(2025)Elyajouri, Abergel, Ysard, Habart, Schirmer,
		Jones, Juvela, \& Tabone}]{elyajouri_jwst_2025}
	Elyajouri, M., Abergel, A., Ysard, N., {et~al.} 2025, A\&A, submitted
	
	\bibitem[{Fitzpatrick \& Massa(1988)}]{fitzpatrick_analysis_1988}
	Fitzpatrick, E.~L. \& Massa, D. 1988, ApJ, 328, 734
	
	\bibitem[{Fitzpatrick {et~al.}(2019)Fitzpatrick, Massa, Gordon, Bohlin, \&
		Clayton}]{fitzpatrick_analysis_2019}
	Fitzpatrick, E.~L., Massa, D., Gordon, K.~D., Bohlin, R., \& Clayton, G.~C.
	2019, ApJ, 886, 108
	
	\bibitem[{Gasman {et~al.}(2023)Gasman, Argyriou, Sloan, Aringer, Álvarez
		Márquez, Fox, Glasse, Glauser, Jones, Justtanont, Kavanagh, Klaassen,
		Labiano, Larson, Law, Mueller, Nayak, Noriega-Crespo, Patapis, Royer, \&
		Vandenbussche}]{gasman_jwst_2023}
	Gasman, D., Argyriou, I., Sloan, G.~C., {et~al.} 2023, A\&A, 673, A102,
	publisher: EDP Sciences
	
	\bibitem[{Godard {et~al.}(2023)Godard, Pineau~des Forêts, Hennebelle, Bellomi,
		\& Valdivia}]{godard_3d_2023}
	Godard, B., Pineau~des Forêts, G., Hennebelle, P., Bellomi, E., \& Valdivia,
	V. 2023, A\&A, 669, A74, aDS Bibcode: 2023A\&A...669A..74G
	
	\bibitem[{Gordon {et~al.}(2009)Gordon, Cartledge, \&
		Clayton}]{gordon_fuse_2009}
	Gordon, K.~D., Cartledge, S., \& Clayton, G.~C. 2009, ApJ, 705, 1320
	
	\bibitem[{Gordon {et~al.}(2023)Gordon, Clayton, Decleir, Fitzpatrick, Massa,
		Misselt, \& Tollerud}]{gordon_one_2023}
	Gordon, K.~D., Clayton, G.~C., Decleir, M., {et~al.} 2023, ApJ, 950, 86
	
	\bibitem[{Gordon {et~al.}(2021)Gordon, Misselt, Bouwman, Clayton, Decleir,
		Hines, Pendleton, Rieke, Smith, \& Whittet}]{gordon_milky_2021}
	Gordon, K.~D., Misselt, K.~A., Bouwman, J., {et~al.} 2021, ApJ, 916, 33
	
	\bibitem[{Gorti \& Hollenbach(2002)}]{gorti_photoevaporation_2002}
	Gorti, U. \& Hollenbach, D. 2002, ApJ, 573, 215, publisher: IOP ADS Bibcode:
	2002ApJ...573..215G
	
	\bibitem[{Habart {et~al.}(2011)Habart, Abergel, Boulanger, Joblin, Verstraete,
		Compiègne, Pineau Des~Forêts, \& Le~Bourlot}]{habart_excitation_2011}
	Habart, E., Abergel, A., Boulanger, F., {et~al.} 2011, A\&A, 527, A122,
	\_eprint: 1012.5324
	
	\bibitem[{Habart {et~al.}(2005)Habart, Abergel, Walmsley, Teyssier, \&
		Pety}]{habart_density_2005}
	Habart, E., Abergel, A., Walmsley, C.~M., Teyssier, D., \& Pety, J. 2005, A\&A,
	437, 177, \_eprint: astro-ph/0501536
	
	\bibitem[{Habart {et~al.}(2004)Habart, Boulanger, Verstraete, Walmsley, \& {G.
			Pineau Des Forêts}}]{habart_empirical_2004}
	Habart, E., Boulanger, F., Verstraete, L., Walmsley, C.~M., \& {G. Pineau Des
		Forêts}. 2004, A\&A, 414, 531
	
	\bibitem[{Habart {et~al.}(2001)Habart, Verstraete, Boulanger, {G. Pineau Des
			Forêts}, Le~Peintre, \& Bernard}]{habart_photoelectric_2001}
	Habart, E., Verstraete, L., Boulanger, F., {et~al.} 2001, A\&A, 373, 702
	
	\bibitem[{Habing(1968)}]{habing_interstellar_1968}
	Habing, H.~J. 1968, Bulletin of the Astronomical Institutes of the Netherlands,
	19, 421, aDS Bibcode: 1968BAN....19..421H
	
	\bibitem[{Hernández-Vera {et~al.}(2023)Hernández-Vera, Guzmán, Goicoechea,
		Maillard, Pety, Petit, Gerin, Bron, Roueff, Abergel, Schirmer, Carpenter,
		Gratier, Gordon, \& Misselt}]{hernandez-vera_extremely_2023}
	Hernández-Vera, C., Guzmán, V.~V., Goicoechea, J.~R., {et~al.} 2023, A\&A,
	677, A152, publisher: EDP Sciences
	
	\bibitem[{Hollenbach \& Tielens(1999)}]{hollenbach_photodissociation_1999}
	Hollenbach, D.~J. \& Tielens, A. G. G.~M. 1999, RMP, 71, 173
	
	\bibitem[{Hwang {et~al.}(2023)Hwang, Pattle, Parsons, Go, \&
		Kim}]{hwang_magnetic_2023}
	Hwang, J., Pattle, K., Parsons, H., Go, M., \& Kim, J. 2023, AJ, 165, 198
	
	\bibitem[{Inoguchi {et~al.}(2020)Inoguchi, Hosokawa, Mineshige, \&
		Kim}]{inoguchi_factories_2020}
	Inoguchi, M., Hosokawa, T., Mineshige, S., \& Kim, J.-G. 2020, MNRAS, 497, 5061
	
	\bibitem[{Kaufman {et~al.}(2006)Kaufman, Wolfire, \&
		Hollenbach}]{kaufman_si_2006}
	Kaufman, M.~J., Wolfire, M.~G., \& Hollenbach, D.~J. 2006, ApJ, 644, 283
	
	\bibitem[{Labiano {et~al.}(2021)Labiano, Argyriou, Álvarez Márquez, Glasse,
		Glauser, Patapis, Law, Brandl, Justtanont, Lahuis, Martínez-Galarza,
		Mueller, Noriega-Crespo, Royer, Shaughnessy, \&
		Vandenbussche}]{labiano_wavelength_2021}
	Labiano, A., Argyriou, I., Álvarez Márquez, J., {et~al.} 2021, A\&A, 656, A57
	
	\bibitem[{Labiano {et~al.}(2016)Labiano, Azzollini, Bailey, Beard, Dicken,
		García-Marín, Geers, Glasse, Glauser, Gordon, Justtanont, Klaassen, Lahuis,
		Law, Morrison, Müller, Rieke, Vandenbussche, \& Wright}]{labiano_miri_2016}
	Labiano, A., Azzollini, R., Bailey, J., {et~al.} 2016, in Observatory
	{Operations}: {Strategies}, {Processes}, and {Systems} {VI}, Vol. 9910
	(SPIE), 947--956
	
	\bibitem[{Law {et~al.}(2023)Law, Morrison, Argyriou, Patapis, Álvarez
		Márquez, Labiano, \& Vandenbussche}]{law_3d_2023}
	Law, D.~R., Morrison, J.~E., Argyriou, I., {et~al.} 2023, AJ, 166, 45,
	publisher: The American Astronomical Society
	
	\bibitem[{Le~Petit {et~al.}(2006)Le~Petit, Nehme, Le~Bourlot, \&
		Roueff}]{le_petit_model_2006}
	Le~Petit, F., Nehme, C., Le~Bourlot, J., \& Roueff, E. 2006, ApJS, 164, 506
	
	\bibitem[{Lefloch \& Lazareff(1994)}]{lefloch_cometary_1994}
	Lefloch, B. \& Lazareff, B. 1994, A\&A, 289, 559, aDS Bibcode:
	1994A\&A...289..559L
	
	\bibitem[{Maillard(2023)}]{maillard_model_2023}
	Maillard, V. 2023, phdthesis, Université Paris sciences et lettres
	
	\bibitem[{Maillard {et~al.}(2021)Maillard, Bron, \&
		Le~Petit}]{maillard_dynamical_2021}
	Maillard, V., Bron, E., \& Le~Petit, F. 2021, A\&A, 656, A65, \_eprint:
	2109.05886
	
	\bibitem[{Misselt {et~al.}(2025)Misselt, Witt, Gordon, Van De~Putte, Trahin,
		Abergel, Noriega-Crespo, Guillard, Zannese, Dell’ova, Baes, Klaassen, \&
		Ysard}]{misselt_jwst_2025}
	Misselt, K., Witt, A.~N., Gordon, K.~D., {et~al.} 2025, A\&A, 700, A158
	
	\bibitem[{Morrison {et~al.}(2023)Morrison, Dicken, Argyriou, Ressler, Gordon,
		Regan, Cracraft, Rieke, Engesser, Alberts, Alvarez-Marquez, Colbert, Fox,
		Gasman, Law, Marin, Gáspár, Guillard, Kendrew, Labiano, Laine,
		Noriega-Crespo, Shivaei, \& Sloan}]{morrison_jwst_2023}
	Morrison, J.~E., Dicken, D., Argyriou, I., {et~al.} 2023, PASP, 135, 075004,
	publisher: The Astronomical Society of the Pacific
	
	\bibitem[{Nakatani \& Yoshida(2019)}]{nakatani_photoevaporation_2019}
	Nakatani, R. \& Yoshida, N. 2019, ApJ, 883, 127, publisher: IOP ADS Bibcode:
	2019ApJ...883..127N
	
	\bibitem[{Neckel \& Sarcander(1985)}]{neckel_spectroscopic_1985}
	Neckel, T. \& Sarcander, M. 1985, A\&A, 147, L1, aDS Bibcode:
	1985A\&A...147L...1N
	
	\bibitem[{Patapis {et~al.}(2024)Patapis, Argyriou, Law, Glauser, Glasse,
		Labiano, Álvarez Márquez, Kavanagh, Gasman, Mueller, Larson, Vandenbussche,
		Lee, Klaassen, Guillard, \& Wright}]{patapis_geometric_2024}
	Patapis, P., Argyriou, I., Law, D.~R., {et~al.} 2024, A\&A, 682, A53
	
	\bibitem[{Pound {et~al.}(2003)Pound, Reipurth, \& Bally}]{pound_looking_2003}
	Pound, M.~W., Reipurth, B., \& Bally, J. 2003, AJ, 125, 2108
	
	\bibitem[{Pound \& Wolfire(2008)}]{pound_photo_2008}
	Pound, M.~W. \& Wolfire, M.~G. 2008, 394, 654, aDS Bibcode: 2008ASPC..394..654P
	
	\bibitem[{Pound \& Wolfire(2011)}]{pound_pdrt_2011}
	Pound, M.~W. \& Wolfire, M.~G. 2011, Astrophysics Source Code Library,
	ascl:1102.022, aDS Bibcode: 2011ascl.soft02022P
	
	\bibitem[{Pound \& Wolfire(2023)}]{pound_photodissociation_2023}
	Pound, M.~W. \& Wolfire, M.~G. 2023, AJ, 165, 25
	
	\bibitem[{Schaerer \& de~Koter(1997)}]{schaerer_combined_1997}
	Schaerer, D. \& de~Koter, A. 1997, A\&A, 322, 598, aDS Bibcode:
	1997A\&A...322..598S
	
	\bibitem[{Schirmer {et~al.}(2021)Schirmer, Habart, Ysard, Bron, Le~Bourlot,
		Verstraete, Abergel, Jones, Roueff, \& Le~Petit}]{schirmer_influence_2021}
	Schirmer, T., Habart, E., Ysard, N., {et~al.} 2021, A\&A, 649, A148
	
	\bibitem[{Sternberg {et~al.}(2014)Sternberg, Le~Petit, Roueff, \&
		Le~Bourlot}]{sternberg_h_2014}
	Sternberg, A., Le~Petit, F., Roueff, E., \& Le~Bourlot, J. 2014, ApJ, 790, 10
	
	\bibitem[{Sternberg \& Neufeld(1999)}]{sternberg_ratio_1999}
	Sternberg, A. \& Neufeld, D.~A. 1999, ApJ, 516, 371, publisher: IOP Publishing
	
	\bibitem[{Störzer \& Hollenbach(1998)}]{storzer_nonequilibrium_1998}
	Störzer, H. \& Hollenbach, D. 1998, ApJ, 495, 853
	
	\bibitem[{Valdivia {et~al.}(2016)Valdivia, Hennebelle, Gérin, \&
		Lesaffre}]{valdivia_h2_2016}
	Valdivia, V., Hennebelle, P., Gérin, M., \& Lesaffre, P. 2016, A\&A, 587, A76,
	aDS Bibcode: 2016A\&A...587A..76V
	
	\bibitem[{Wakelam {et~al.}(2017)Wakelam, Bron, Cazaux, Dulieu, Gry, Guillard,
		Habart, Hornekær, Morisset, Nyman, Pirronello, Price, Valdivia, Vidali, \&
		Watanabe}]{wakelam_h_2017}
	Wakelam, V., Bron, E., Cazaux, S., {et~al.} 2017, Molecular Astrophysics, 9, 1,
	\_eprint: 1711.10568
	
	\bibitem[{Warren \& Hesser(1977)}]{warren_photometric_1977}
	Warren, Jr., W.~H. \& Hesser, J.~E. 1977, ApJS, 34, 115
	
	\bibitem[{Weingartner \& Draine(2001)}]{weingartner_dust_2001}
	Weingartner, J. \& Draine, B. 2001, ApJ, 548, 296
	
	\bibitem[{Wolfire {et~al.}(2022)Wolfire, Vallini, \&
		Chevance}]{wolfire_photodissociation_2022}
	Wolfire, M.~G., Vallini, L., \& Chevance, M. 2022, ARAA, 60, 247
	
	\bibitem[{Wright {et~al.}(2023)Wright, Rieke, Glasse, Ressler, Marín, Aguilar,
		Alberts, Álvarez Márquez, Argyriou, Banks, Baudoz, Boccaletti, Bouchet,
		Bouwman, Brandl, Breda, Bright, Cale, Colina, Cossou, Coulais, Cracraft,
		Meester, Dicken, Engesser, Etxaluze, Fox, Friedman, Fu, Gasman, Gáspár,
		Gastaud, Geers, Glauser, Gordon, Greene, Greve, Grundy, Güdel, Guillard,
		Haderlein, Hashimoto, Henning, Hines, Holler, Detre, Jahromi, James, Jones,
		Justtanont, Kavanagh, Kendrew, Klaassen, Krause, Labiano, Lagage, Lambros,
		Larson, Law, Lee, Libralato, Alverez, Meixner, Morrison, Mueller, Murray,
		Mycroft, Myers, Nayak, Naylor, Nickson, Noriega-Crespo, Östlin,
		O’Sullivan, Ottens, Patapis, Penanen, Pietraszkiewicz, Ray, Regan,
		Roteliuk, Royer, Samara-Ratna, Samuelson, Sargent, Scheithauer, Schneider,
		Schreiber, Shaughnessy, Sheehan, Shivaei, Sloan, Tamas, Teague, Temim,
		Tikkanen, Tustain, Dishoeck, Vandenbussche, Weilert, Whitehouse, \&
		Wolff}]{wright_mid-infrared_2023}
	Wright, G.~S., Rieke, G.~H., Glasse, A., {et~al.} 2023, PASP, 135, 048003,
	publisher: The Astronomical Society of the Pacific
	
	\bibitem[{Álvarez Márquez {et~al.}(2023)Álvarez Márquez, Labiano, Guillard,
		Dicken, Argyriou, Patapis, Law, Kavanagh, Larson, Gasman, Mueller, Alberts,
		Brandl, Colina, García-Marín, Jones, Noriega-Crespo, Shivaei, Temim, \&
		Wright}]{alvarez-marquez_nuclear_2023}
	Álvarez Márquez, J., Labiano, A., Guillard, P., {et~al.} 2023, A\&A, 672,
	A108
	
\end{thebibliography}
\end{document}